\documentclass[journal]{IEEEtran}
\usepackage{amsmath,epsfig,color}
\usepackage{url}
\usepackage{bigstrut,multirow,rotating}
\usepackage{textcomp,booktabs}
\usepackage{tabularx}
\usepackage{floatrow}
\newfloatcommand{capbtabbox}{table}[][\FBwidth]
\usepackage{caption}
\usepackage{colortbl}
\usepackage{xcolor}
\usepackage{multirow}
\usepackage{booktabs}
\usepackage{underscore}
\usepackage[marginal]{footmisc}
\usepackage{subfigure}
\usepackage{algorithm,algorithmic}
\usepackage{amsfonts}

\begin{document}
\newcommand{\bl}[1]{{\color{black}#1}}
\newcommand{\bbl}[1]{{\color{black}#1}}
\title{Learning Generalized Spatial-Temporal Deep Feature Representation for No-Reference Video Quality Assessment}
\author{Baoliang Chen, Lingyu Zhu, Guo Li, Fangbo Lu, Hongfei Fan and Shiqi Wang~\IEEEmembership{Member,~IEEE}

\thanks{This research was supported in part by the National Natural Science Foundation of China under 62022002, in part by the Hong Kong Research Grants Council, Early Career Scheme (RGC ECS) under Grant 21211018, General Research Fund (GRF) under Grant 11203220, in part by the City University of Hong Kong, Teaching Development Grants (CITYU TDG) Grant 6000713 and Applied Research Grant (ARG) grant 9667192.}

\thanks{B. Chen, L. Zhu and S. Wang are with the Department of Computer Science, City University of Hong Kong, Hong Kong (e-mail: blchen6-c@my.cityu.edu.hk; lingyzhu@cityu.edu.hk; shiqwang@cityu.edu.hk). G. Li, F. Lu and H. Fan are with Kingsoft Cloud, China (email: liguo136009@foxmail.com; lufangbo@gmail.com; fanhongfei@kingsoft.com). Corresponding author: Shiqi Wang.}

}

\maketitle

\begin{abstract}


In this work, we propose a no-reference video quality assessment method, aiming to achieve  high-generalization capability in cross-content, -resolution and -frame rate quality prediction. In particular, we evaluate the quality of a video by learning effective feature representations in spatial-temporal domain. In the spatial domain, to tackle the resolution and content variations, 
we impose the Gaussian distribution constraints on the quality features. The unified distribution can significantly reduce the domain gap between different video samples, resulting in  more generalized quality feature representation. Along the temporal dimension, inspired by the mechanism of visual perception, we propose a pyramid temporal aggregation module by involving the short-term and long-term memory to aggregate the frame-level quality. Experiments show that our method  outperforms the state-of-the-art methods on cross-dataset settings, and achieves comparable performance on intra-dataset configurations, demonstrating the high-generalization capability of the proposed method. 
\bl{The codes are released at \textit{\url{https://github.com/Baoliang93/GSTVQA}}.} 

\end{abstract}

\IEEEpeerreviewmaketitle

\section{Introduction}
\IEEEPARstart{T}{here} has been an increasing demand for accurately predicting the quality of videos, coinciding with the exponentially growing of video data. In the context of video big data, it becomes extremely difficult and costly to rely solely on human visual system to conduct timely quality assessment. As such, objective video quality assessment (VQA), the goal of which is to design computational models that automatically and accurately predict the perceived quality of videos, has become more prominent. According to the application scenarios regarding the availability of the pristine reference video, the assessment of video quality can be categorized into full-reference VQA (FR-VQA), reduced-reference VQA (RR-VQA) and no-reference VQA (NR-VQA). Despite remarkable progress, the NR-VQA of real-world videos, which has received great interest due to its high practical utility, is still very challenging especially when the videos are acquired, processed and compressed with diverse devices, environments and algorithms.

For NR-VQA, numerous methods have been proposed in the literature, and the majority of them rely on a machine learning pipeline based on the training of quality prediction model with labeled data. Methods relying on  handcrafted features  \cite{tu2020ugc,mittal2012no,zhang2015feature,mittal2015completely} and deep learning features \cite{liu2018end, zhu2020metaiqa, fang2020perceptual, ren2017ran4iqa} have been developed, with the assumption that the training and testing data are drawn from closely aligned feature spaces. However, it is widely acknowledged that different distributions of training and testing data create the risk of poor generalization capability, and as a consequence, inaccurate predictions could be obtained on the videos that hold dramatically different statistics compared to those in the training set. 
The underlying design principle of the proposed VQA method is learning features with high generalization capability, such that the model is able to deliver high quality prediction accuracy of videos that are not sampled from the domain of the training data. This well aligns real application scenarios when the testing data are unknown. To verify the performance of our method, we conduct experiments on four cross-dataset settings with  available databases, including KoNViD-1k \cite{hosu2017konstanz}, LIVE-Qualcomm \cite{ghadiyaram2017capture}, LIVE-VQC \cite{sinno2018large} and CVD2014 \cite{nuutinen2016cvd2014}. Experimental results have demonstrated superior performance of our method over existing state-of-the-art models with a significant margin. The main contributions of this paper are as follows,
\begin{itemize}
\item We propose an objective NR-VQA model that is capable of automatically accessing the perceptual quality of videos resulting from different acquisition, processing and compression techniques. The proposed model is driven by learning features that specifically characterize the quality, and is able to deliver high prediction accuracy for videos that hold dramatically different characteristics compared to the training data. 

\item In the spatial domain, we develop a multi-scale feature extraction scheme to explore the quality features in different scales, and an attention module is further incorporated to adaptively weight the features by their importance. We further unify the quality features of each frame with a Gaussian distribution where the mean and variance of the distribution are learnable. As such, the domain gap of different video samples caused by the content and distortion types can be further reduced by such a normalization operation.

\item In the temporal domain, a pyramid temporal pooling layer is proposed to account for the quality aggregation in temporal domain. The pyramid temporal pooling can make temporal pooling independent of the number of frames of the input video and aggregate the short-term and long-term quality levels of a video in a pyramid manner, which further enhances the generalization ability of the proposed model.
\end{itemize}

\section{Related Works}

\subsection{No-reference Image Quality Assessment} 
Generally speaking, general purpose no-reference image quality assessment (NR-IQA) methods, which do not require any prior information of distortion types, hold the assumption that the destruction of
``naturalness'' could be the useful clue in quality assessment. The so-called natural scene statistic (NSS) approaches rely on a series of handcrafted features extracted in both spatial and frequency domains. Mittal \textit{et al.} \cite{mittal2012no} investigated NSS features by exploiting the local spatial normalized luminance coefficients.  Xue \textit{et al.} \cite{xue2014blind} combined the gradient magnitude (GM) and Laplacian of Gaussian (LoG) features together, and the results show that joint statistics GM-LoG could obtain desirable performance for NR-IQA task. Gu \textit{et al.} \cite{gu2014using} proposed a general purpose NR-IQA metric by exploiting the features that are highly correlated to human perception, including structural information and gradient magnitude. The Distortion Identification-based Image Verity and Integrity Evaluation (DIIVINE) method was developed by Moorthy \textit{et al.} \cite{moorthy2011blind} with a two-stage framework, which includes distortion identification and support vector regression (SVR) to quality scores for distorted natural images. Narwaria \textit{et al.} quantified structural representation in images with the assistant of singular value decomposition (SVD), and formulated  quality prediction as a regression problem to predict image score using SVR. \bl{Instead of using SVR for quality regression, in \cite{camps2018one}, based on a one-class regressor, the Gaussian Process (GP) is utilized for  quality estimation of transperineal ultrasound images. However, only binary label generated may not be sufficient for natural image quality assessment. The GP is also used in \cite{tang2014blind, wu2017blind}. In \cite{tang2014blind}, a deep belief network with a nonlinear kernel regression function is introduced to formulate a GP kernel for NR-IQA. In  \cite{wu2017blind}, an uncertainty-aware evaluator is created with the GP. In particular, the perceptually consistent neighbors of a test image are selected by  kernel density estimation (KDE). Comparing with those GP based methods, the Gaussian distribution utilized in our method is dedicated to global quality features regularization, such that the quality score can be estimated by the distribution.} Another efficient NR-IQA method in \cite{saad2012blind} explored the discrete cosine transform (DCT) domain statistics to predict perceptual quality. Zhang \textit{et al.} \cite{zhang2013no} designed the  DErivative Statistics-based Image QUality Evaluator (DESIQUE), exploiting statistical features related to quality in spatial and frequency domains, which can be fitted by a generalized  Gaussian distribution model to estimate image quality. Just noticeable difference (JND) model are an another human perception inspired method for IQA and was also widely studied in the literature \textit{e.g.} \cite{chen2019asymmetric, kim2017no,gu2019multiple}. In stead of feature extraction for quality regression, a label transfer strategy was adopted in \cite{wu2015blind} where the quality score of an image can be predicted based on the assumption that similar images share similar perceptual qualities.

Recently, sophisticated deep learning based NR-IQA methods have been developed, demonstrating superior prediction performance over traditional methods. Zhang \textit{et al.} \cite{zhang2018blind} proposed a deep bilinear model for NR-IQA that is suitable for the quality assessment of synthetic and real distorted images. The bilinear model includes two convolutional neural networks (CNNs){:} S-CNN and pre-trained VGG, which account for the synthetic and real-world distortions, respectively. 
In view of the challenges in cross-distortion-scenario prediction, Zhang \textit{et al.} \cite{zhang2020learning} used massive image pairs composed of multiple databases simultaneously to train a unified blind image quality assessment model. The Neural IMage Assessment (NIMA) model \cite{talebi2018nima} which tackles the problem of understanding visual aesthetics was trained on large-scale Aesthetic Visual Analysis (AVA) dataset \cite{murray2012ava} to predict the distribution of quality ratings. Su \textit{et al.} \cite{su2020blindly} proposed an adaptive multi-scale hyper-network architecture, which consists of two modules{:} content understanding and quality prediction networks, to predict quality score based on captured local and global distortions. Zhu \textit{et al.} \cite{zhu2020metaiqa} developed a reference-free IQA metric based on deep meta-learning, which can be easily adapted to unknown distortions by learning meta-knowledge shared by human.  Bosse \textit{et al.} \cite{bosse2017deep} proposed a data-driven end-to-end method for FR and NR image quality assessment task simultaneously. \bl{}


\subsection{No-reference Video Quality Assessment}
 Recently, considerable efforts have been dedicated to VQA, in particular for quantifying the compression and transmission artifacts. Manasa \textit{et al.} \cite{manasa2016optical} developed the NR-VQA model based on the statistics of optical flow. In particular, to capture the influence of distortion on optical flow, statistical irregularities of optical flow at patch level and frame level are quantified, which are further combined with the SVR to predict the perceptual video quality. Li \textit{et al.} \cite{li2015no} developed an NR-VQA by combining 3D shearlet transform and deep learning to pool the quality score. Video Multi-task End-to-end Optimized neural Network (V-MEON) \cite{liu2018end} is an NR-VQA technique designed based on feature extraction with 3D convolutional layer. Such spatial-temporal features could lead to better quality prediction performance. Korhonen \textit{et al.} \cite{korhonen2019two} extracted Low Complexity Features (LCF) from full video sequences and High Complexity Features (HCF) from key frames, following which SVR is used to predict video score. Vega \textit{et al.} \cite{vega2017deep} focused on packet loss effects for video streaming settings, and an unsupervised learning based model is employed at the video server (offline) and the client (in real-time). In ~\cite{li2019quality}, Li \textit{et al.} integrated both content and temporal-memory in the NR-VQA model, and the gated recurrent unit (GRU) is used for long-term temporal feature extraction. \bbl{To incorporate the motion perception in VQA task, a Recurrent-In-Recurrent Network (RIRNet) was proposed in \cite{chen2020rirnet}. In RIRNet, the motion information derived from different temporal frequencies can be fused efficiently.} You \textit{et al.} \cite{you2019deep} used 3D convolution network to extract local spatial-temporal features from small clips in the video. This not only addresses the problem of insufficient training data, but also effectively captures the perceptual quality features which are finally fed into the LSTM  network to predict the perceived video quality.

\begin{figure*}[ht]
\begin{center}
\includegraphics[width=1.0\textwidth]{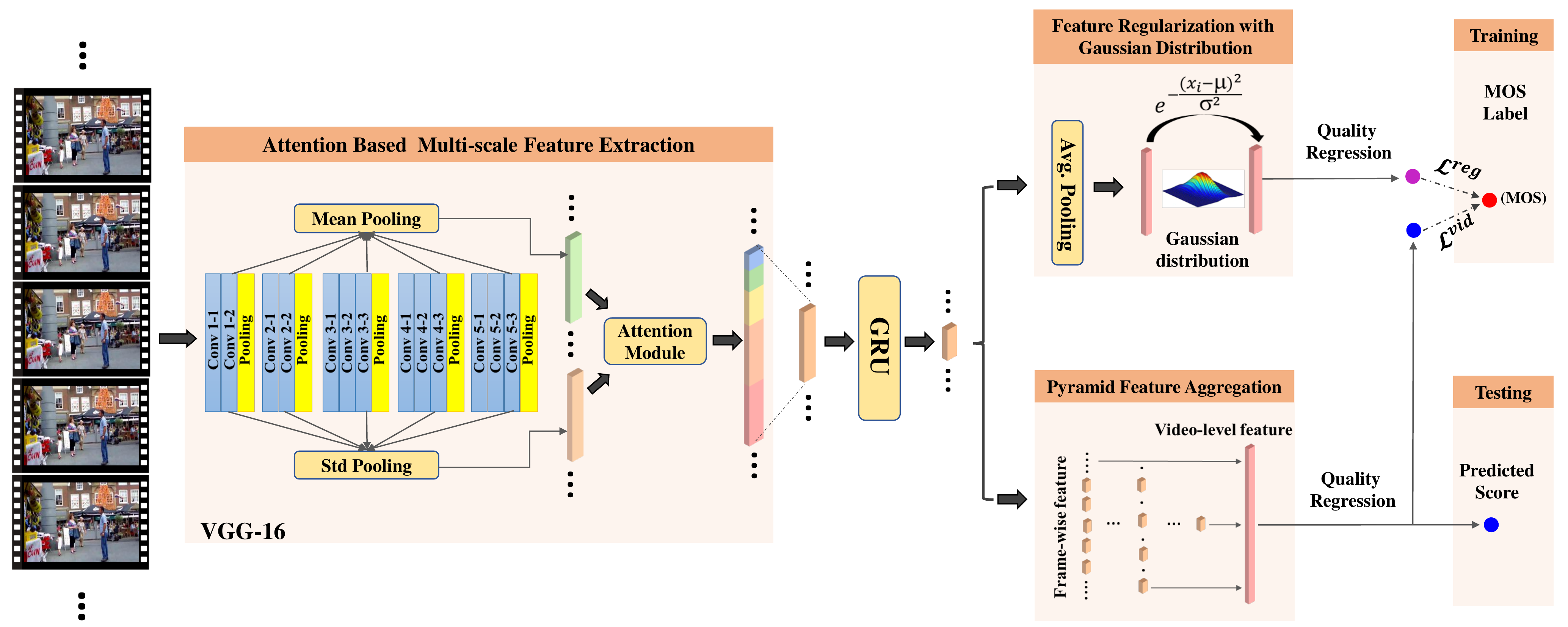}
\caption{\bl{The framework of the proposed generalized NR-VQA model. For each frame of the input video, we first utilize the pre-trained VGG16 network to extract the multi-scale features with an attention module. Subsequently, the extracted features are further processed by a fully connected layer to reduce its dimension, followed by a GRU module to acquire the frame-level quality features. We further regularize the frame-level quality features by forcing the features to obey Gaussian distributions via adversarial learning. Finally, a pyramid pooling strategy is utilized for temporal quality aggregation inspired by short-term and long-term memory effects.}}
\label{fig:flow}
\end{center}
\end{figure*}

\subsection{Domain Generalization} 
The VQA problem also suffers from the domain gap between the labeled training data (source domain) and unseen testing data (target domain), leading to the difficulty that the trained model in the labeled data cannot generalize well on the unseen data. These feature gaps may originate from different resolutions, scenes, acquisition devices/conditions and processing/compression artifacts. Over the past years, numerous attempts have been made to address domain generalization problem by learning domain-invariant representations~\cite{muandet2013domain, erfani2016robust, xie2017controllable,ghifary2015domain, xu2014exploiting,motiian2017unified}, which lead to promising results. In \cite{andrew2013deep}, Canonical Correlation Analysis (CCA) was proposed to learn the shareable information among domains.  Muandet \textit{et al.} \cite{muandet2013domain} proposed to leverage Domain Invariant Component Analysis (DICA) to minimize the distribution mismatch across domains.  In \cite{carlucci2019domain} Carlucci \textit{et al.} learn the  generalized representation by shuffling the image patches and this idea was further extended by \cite{wang2020heterogeneous}, in which the samples across multiple source domains are mixed for heterogeneous domain generalization task. The generalization of adversarial training \cite{goodfellow2014explaining, sinha2017certifiable} has also been extensively studied. For example, Li \textit{et al.} \cite{li2018domain} proposed the MMD-AAE model which extends adversarial autoencoders by imposing the Maximum Mean Discrepancy (MMD) measure to align the distributions among different domains. Instead of training domain classifiers in our work due to sample complexity \cite{schmidt2018adversarially} and uncontrolled conditions (scenes, distortion types, motion, resolutions, \textit{etc.}), we further regularize the learned feature to follow Gaussian distribution via adversarial training, 
shrinking the learned feature mismatch across domains. 

\section{The Proposed Scheme}
We aim to learn an NR-VQA model with high generalization capability for real-world applications. Generally speaking, three intrinsic attributes that govern the generalization capability of VQA are considered, including spatial resolution, frame rate and video content (e.g., the captured scenes and the distortion type). As shown in Fig.~\ref{fig:flow}, we first extract the frame-level quality features with a pretrained VGG16 model~\cite{simonyan2014very}, inspired by the repeatedly proven evidence that such features could reliably reflect the visual quality~\cite{zhang2018blind,ding2020image, li2019quality,li2017image}. To encode the generalization capability to different spatial resolutions into feature representation, statistical pooling moments are leveraged and the features in the five convolution stages (from top layer to bottom layer) are aggregated with the channel attention. To further enhance the generalization capability to unseen domains, the large distribution gap between the source and target domains are blindly compensated by regularizing the learned quality feature into a unified distribution. In the temporal domain, a pyramid aggregation module is further proposed, leading to the final quality features for quality prediction.

\subsection{Attention Based Multi-scale Feature Extraction}
\begin{figure}[t]
\begin{minipage}[b]{0.95\linewidth}
  \centering
  \centerline{\includegraphics[width=1\linewidth]{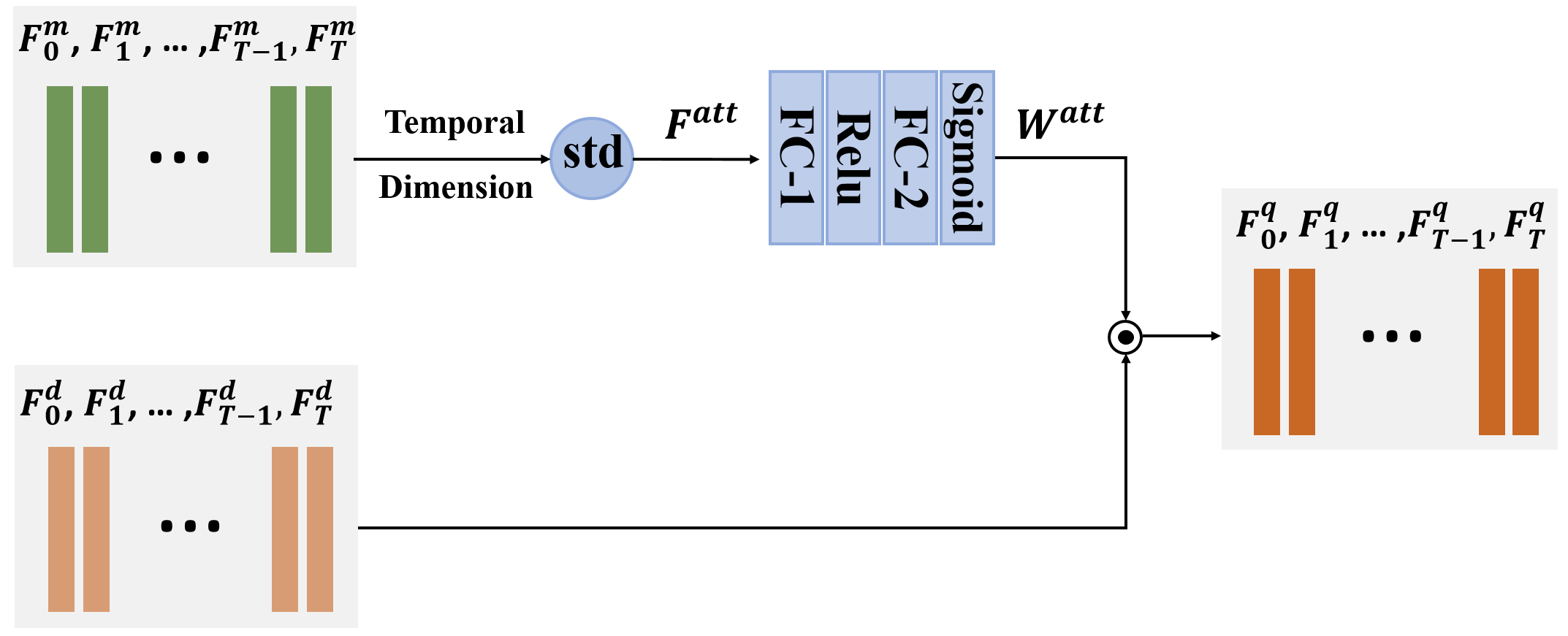}}
\end{minipage}
\caption{Illustration of the attention module for feature extraction.}
\label{fig:att}
\end{figure}

Herein, the feature representation that is equipped with strong generalization capability in terms of the spatial resolution of a single frame is obtained based on the pretrained VGG  ConvNets. It is widely acknowledged that the pooling moments determine the discriminability of features, and we adopt the widely acknowledged mean and standard deviation (std) based pooling strategies. 
In particular, for frame $i$, supposing the mean pooling and std pooling results of the output feature at stage $s$ $(s \in \{1,2,3,4,5\}) $ as $\textbf{\textit{M}}_i^s$ and $\textbf{\textit{D}}_i^s$ respectively, the multi-scale quality representations can be acquired by concatenating the pooled features at each stage as follows,
\begin{equation}\label{mstd1}
\begin{array}{l}
\text { $\textbf{\textit{{F}}}_{i}^{m}$}= Concat(\textbf{\textit{M}}_{i}^{1},  \textbf{\textit{M}}_{i}^{2},...,\textbf{\textit{M}}_{i}^{5}), \\\\
\text { $\textbf{\textit{{F}}}_{i}^{d}$}= Concat(\textbf{\textit{D}}_{i}^{1},  \textbf{\textit{D}}_{i}^{2},...,\textbf{\textit{D}}_{i}^{5}),
\end{array}
\end{equation}
where $\textbf{\textit{F}}_{i}^{m}$ and $\textbf{\textit{F}}_{i}^{d}$ stand for the multi-scale mean feature and std feature of frame $i$. However, it may not be feasible to concatenate the two pooled features straightforwardly for quality regression, due to the high relevance of $\textbf{\textit{F}}_{i}^{m}$ with the semantic information~\cite{wan2019information}. As a result, the learned model tends to overfit to the specific scenes in the training set. Here, instead of discarding the $\textbf{\textit{F}}_{i}^{m}$, as shown in Fig.~\ref{fig:att}, the $\textbf{\textit{F}}_{i}^{m}$ is regarded as the semantically meaningful feature working as the integral part in the attention based multi-scale feature extraction. To be specific, for $T$ frames, given $[\textbf{\textit{F}}_{1}^{m},\textbf{\textit{F}}_{2}^{m},..., \textbf{\textit{F}}_{T-1}^{m},\textbf{\textit{F}}_{T}^{m}]$, we first calculate the std of each channel  along the temporal dimension as follows, 
\begin{equation}\label{mstd2}
\text { $\textbf{\textit{F}}^{att}$ }= \sqrt{\frac{1}{T-1} \sum_{i=1}^{T}(\textbf{\textit{F}}_{i}^{m}-\overline{\textbf{\textit{F}}}^{m})^{2}},\\\\
\end{equation}
and 
\begin{equation}\label{mstd3}
\overline{\textbf{\textit{F}}}^{m}=\frac{1}{T} \sum_{i=1}^{T} \textbf{\textit{F}}_{i}^{m},
\end{equation}
where the frame index is denoted as \(i\). Given \(\textbf{\textit{F}}^{att}\), two fully connected layers are learned to implement the attention mechanism, as shown in Fig.~\ref{fig:att}, 
\begin{equation}\label{mstd4}
\text { $\textbf{\textit{W}}^{att}$ }= Sigmoid(FC_2(ReLu(FC_1(\textbf{\textit{F}}^{att})))),
\end{equation}
where $FC_1(\cdot)$ and $FC_2(\cdot)$ represent the two fully connected layers. 
The underlying principle is the attention weight in each channel depends on the corresponding variance along the temporal domain, which is highly relevant with the video content variations. As such, such nested pooling with spatial mean and temporal std could provide the attention map by progressively encoding the spatial and temporal variations into a global descriptor. Then the frame-specific quality representation $\textbf{\textit{F}}_{i}^{q}$ 
can be obtained by $\textbf{\textit{{F}}}_{i}^{d}$ and its attention weight $\textbf{\textit{W}}^{att}$ as follows,
\begin{equation}\label{mstd5}
\textbf{\textit{{F}}}_{i}^{q}= \textbf{\textit{W}}^{att} \odot \textbf{\textit{{F}}}_{i}^{d},
\end{equation}
where the ``$\odot$'' represents the element wise multiplication.

\subsection{Feature Regularization with Gaussian Distribution}
Given the frame-level quality feature $\textbf{\textit{F}}_{i}^{q}$, the Gated Recurrent Unit (GRU) \cite{cho2014learning} layer is utilized to refine the frame-level feature by involving the temporal information. In particular, we use a fully connected layer (denoted as $FC_3$) to reduce the redundancies of VGG feature, following which the resultant feature is processed by a GRU layer,
\begin{equation}\label{mstd0}
\text { $\textbf{\textit{F}}_{i}^{gru}$ }= GRU(FC_3(\textbf{\textit{F}}_{i}^{q})).
\end{equation}
However, we argue that the $\textbf{\textit{F}}_{i}^{gru}$ is still not generalized enough to different scenes and distortion types.
To enhance the generalization capability of $\textbf{\textit{F}}_{i}^{gru}$, we resort to feature regularization, expecting to learn the quality feature with a unified distribution. The underlying assumption of generalizing to an unseen domain is that there exists a discrete attribute separating the data into different domains. However, a naïve extension to VQA may be confused by numerous discrete or continuous attributes (e.g., scene, distortion type, motion, resolution) for domain classification. As such, instead of dividing the data into different domains, we restrict the frame-level feature subject to a mixture Gaussian distribution by a GAN based model, and moreover the mean and variance of the presumed Gaussian distribution can also be adaptively learned. To be specific, as shown in Fig.~\ref{fig:flow}, we first average the extracted $\textbf{\textit{F}}^{gru}$ of each frame as follows,
\begin{equation}\label{Eqn:favg}
\textbf{\textit{F}}^{avg} = \frac{1}{T}\sum_{i=1}^{T}\textbf{\textit{F}}_{i}^{gru}.
\end{equation}
Herein, we treat the feature $\textbf{\textit{F}}^{avg}$ extractor as the generator \(G(\cdot)\) of a GAN model and we sample the same dimension vector (denoted as $\textbf{\textit{F}}^{gaus}$) from the  prior Gaussian distribution as reference. Then the discriminator \(D(\cdot)\) tries to distinguish the generated feature from the sampled vector. The GAN model is trained through the following adversarial loss,
\begin{equation}\label{Eqn:gan}
 \min _{G} \max _{D} E_{\mathbf{z} \sim g(\mathbf{z})}[\log D(\mathbf{z})]+E_{\mathbf{x} \sim q(\mathbf{x})}[\log (1-D(G(\mathbf{x})))],
\end{equation}
where $\mathbf{z}$ is the vector $\textbf{\textit{F}}^{gaus}$ sampled from  Gaussian distribution $g(x)$, $\mathbf{x}$ is the input video and $G(\mathbf{x})$ generates the feature $\textbf{\textit{F}}^{avg}$. When the network is trained in the first $N$ epochs, we constrain the $g(x)$ to be the standard Gaussian distribution with mean \textit{$\mu=0$} and variance \textit{$\sigma=1$}. However, this imposes a strong constraint that the features in each dimension share the Gaussian distribution with identical mean and variance. Generally speaking, each dimension of the feature is expected to represent a perceptual relevance attribute for quality inference, such that they ultimately follow different Gaussian distributions parameterized by different \textit{$\mu$} and \textit{$\sigma$}. 
This motivates us to adapt the mean and variance of prior Gaussian distribution of each dimension via learning. More specifically, to learn the parameters $\boldsymbol{\mu} = [\mu_1,\mu_2,...,\mu_L]$ and $\boldsymbol{\sigma}  = [\sigma_1,\sigma_2,...,\sigma_L]$ where \(L\) is the dimension of $\textbf{\textit{F}}^{avg}$, we impose the constraint on 
$\textbf{\textit{F}}^{avg}$ to regress the quality score
\begin{equation}\label{frame}
Q^{reg} = \frac{1}{L} \sum_{l=1}^{L} e^{-\frac{\left(\textbf{\textit{F}}^{avg}(l)-\boldsymbol{\mu}(l)\right)^{2}}{\boldsymbol{\sigma}(l)^{2}}}.
\end{equation}
Here, we use $Q^{reg}$ to represent the predicted quality score of the input video, and we aim to regress $Q^{reg}$ towards the ground-truth mean opinion score (MOS) via learning the optimal $\boldsymbol{\mu}$ and $\boldsymbol{\sigma}$. Moreover, \(l\) indicates the $l-th$ dimension. During the training of the network, after every $N$ epochs, we  use the Gaussian distribution  with learned $\boldsymbol{\mu}$ and $\boldsymbol{\sigma}$ to replace the distribution in previous $N$ epochs. From the experimental results, we also find such an adaptive refreshing mechanism can further improve the performance of our model compared with standard Gaussian distribution.

\subsection{Pyramid Feature Aggregation}
\begin{figure}[t]
\begin{minipage}[b]{1.0\linewidth}
  \centering
  \centerline{\includegraphics[width=1\linewidth]{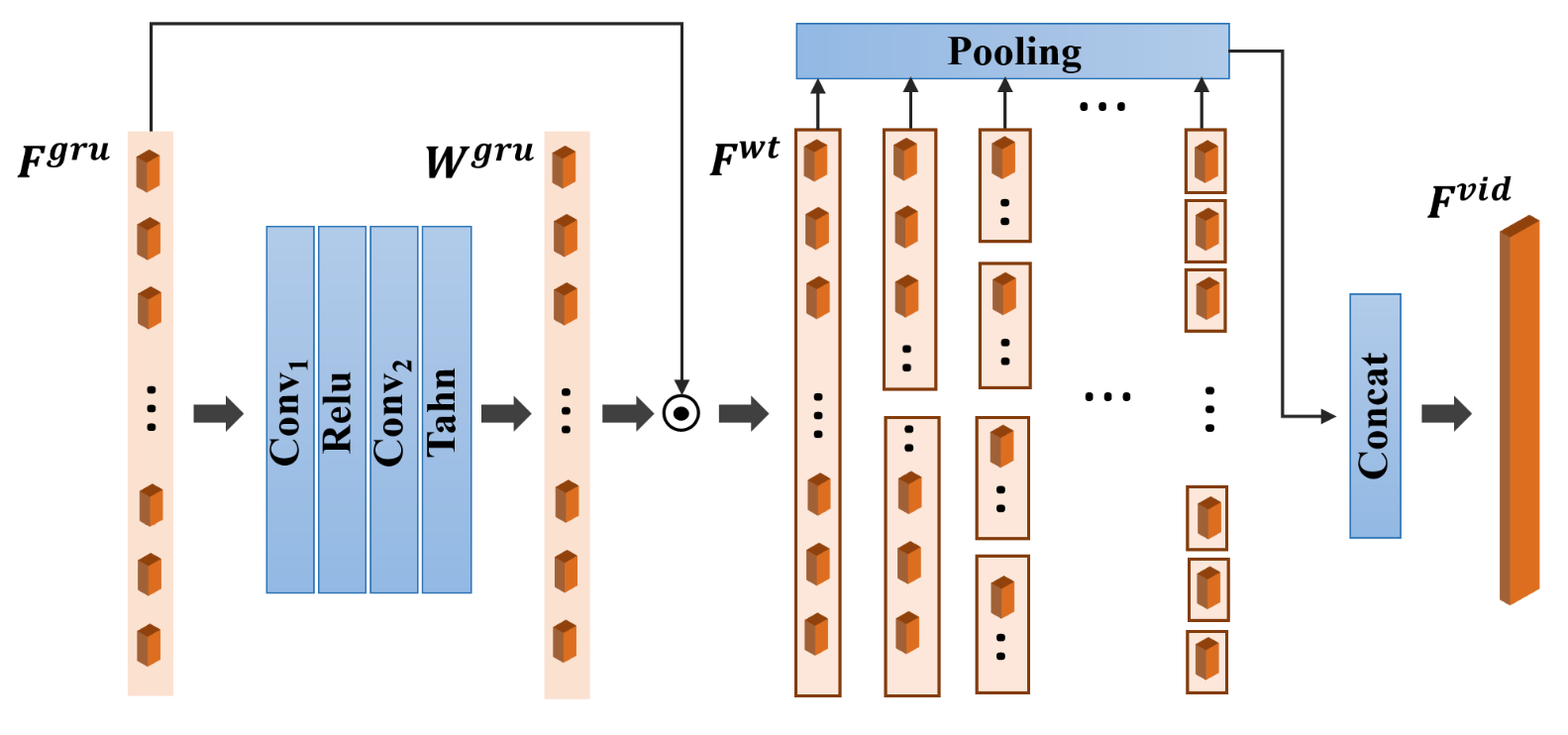}}
\end{minipage}
\caption{Illustration of PFA module.}
\label{fig:pfa}
\end{figure}

Temporal domain aggregation plays an indispensable role in objective VQA models. We consider two cognitive mechanisms in visual quality perception~\cite{hochreiter1997long,zhang2016video}. 
The short-term memory effect persuades us to consider the video quality for each localized time-frame, due to the consensus that subjects are resistant in their opinion and prefer consistent quality when watching the video. Moreover, the long-term memory effect suggests that the global pooling over the whole video sequence in a coarse-to-fine manner could lead to the final video quality. Therefore, we imitate such perception mechanisms with a pyramid feature aggregation (PFA) strategy. In the PFA, the short-term memory and long-term memory are incorporated and the aggregation result is independent of the number of frames. More specifically, as illustrated in Fig.~\ref{fig:pfa}, in the bottom layer of the pyramid, for $\textbf{\textit{F}}^{gru}$, we calculate its weight $\textbf{\textit{W}}^{gru}$ by synthesizing it with its surrounding \(k\) frames,
\begin{equation}\label{cov1d}
\textbf{\textit{W}}^{gru} = Tahn(Conv_2(ReLu(Conv_1(\textbf{\textit{F}}^{gru})))),
\end{equation} 
where the $Conv_1(\cdot)$ and  $Conv_2(\cdot)$ are two 1D-CNNs and their kernel sizes are all set to $2k+1$. Moreover, $Tahn(\cdot)$ and $Relu(\cdot)$ are the activation functions, and $Tahn(\cdot)$ is defined as follows,
\begin{equation}
Tahn(u)=\frac{\left(e^{u}-e^{-u}\right)}{\left(e^{u}+e^{-u}\right)}.
\end{equation}
Then the weighted frame-level quality feature $\textbf{\textit{F}}^{wt}$ can be acquired,
\begin{equation}\label{pym}
\textbf{\textit{F}}^{wt} = \textbf{\textit{W}}^{gru} \odot \textbf{\textit{F}}^{gru}.
\end{equation} 
Subsequently, the weighted frame-level features along the temporal dimension are aggregated in a pyramid manner. In general, the perceivability along the temporal dimension determines the sampling density governed by the number of layers. Herein, we empirically set the number of layers with a constant number 7. To be specific, for the $m-th$ layer ($m \in\{1,2,3...,7\}$), the weighted frame-level features are aggregated into a vector with the dimension $h \times 2^{m-1}$, where \(h\) denotes the feature dimension in $\textbf{\textit{F}}^{gru}_i$. In other words, the video is averagely divided into $2^{m-1}$ time slots, and within each time slot, average feature pooling is performed for aggregation. Finally, we concatenate the aggregated features of all layers, leading to the video-level quality feature with a constant dimension that is independent of the number of frames and frame rate, $\textbf{\textit{F}}^{vid} \in \mathbb{R}^{h \times (2^{m}-1)}$. We first apply a fully connected layer ($FC_4$) to reduce the channel from $h$ to 1, then another fully connected layer ($FC_5$) is adopted to synthesize the pyramid aggregated features. As such, the quality of input videos can be predicted as follows,
\begin{equation}\label{vid}
\text{${Q^{vid}}$}= FC_5(FC_4(\textbf{\textit{F}}^{vid})),
\end{equation}
where  ${Q^{vid}}$ is the prediction score. This strategy provides more flexibility than single layer aggregation by incorporating the variations along the temporal dimension.  

\subsection{Objective Loss Function}
The final loss function involves the frame-level and video-level quality regression results acquired in  Eqn.~\eqref{frame} and Eqn.~\eqref{vid}, as well as the distribution based feature regularization,
\begin{equation}
\min _{G, Q^{vid},Q^{reg}} \max _{D} \mathcal{L}^{\mathrm{vid}}+\lambda_{1} \mathcal{L}^{\mathrm{reg}}+\lambda_{2} \mathcal{R}^{\mathrm{gan}}
\label{Eqn:loss}
\end{equation}
where
\begin{equation}
\begin{array}{c}
\mathcal{L}^{\mathrm{vid}}=\left|Q^{vid}-M O S\right|, \\\\
\mathcal{L}^{\mathrm{reg}}=\left|Q^{reg}-M O S\right|, \\\\
\mathcal{R}^{\mathrm{gan}}=E_{\mathbf{z} \sim g(\mathbf{z})}[\log D(\mathbf{z})]+E_{\mathbf{x} \sim q(\mathbf{x})}[\log (1-D(G(\mathbf{x})))].
\end{array}
\end{equation}
Herein, $\lambda_{1}$ and $\lambda_{2}$  are two trade-off parameters. In the testing phase, we use the $Q^{vid}$ as the final quality score that our model predicts.

\section{Experimental Results}

\subsection{Experimental Setups}
\subsubsection{Datasets}
\bl{various video databases have been proposed for VQA, including BVI-HD \cite{zhang2018bvi}, MCL-JCV \cite{wang2016mcl}, MCML-4K \cite{cheon2017subjective}, LIVE-Netflix \cite{bampis2017study}, KoNViD-1k \cite{hosu2017konstanz}, LIVE-Qualcomm \cite{ghadiyaram2017capture}, LIVE-VQC \cite{sinno2018large} and CVD2014 \cite{nuutinen2016cvd2014}. The BVI-HD VQA database is proposed for compressed video  quality assessment. In MCL-JCV, a H.264/AVC encoded video dataset is constructed with a new just-noticeable difference (JND) based VQA metric. In MCML-4K dataset,  the 4K ultra-high-definition (UHD) videos are compressed by three video coding techniques. LIVE-Netflix is created  to simulate a typical video streaming application, using long  video sequences and  Netflix content. However, the above VQA databases are all constructed for compressed videos. To validate the proposed NR-VQA method on the videos in the wild, we evaluate our model on four popular VQA databases, including  CVD2014 \cite{nuutinen2016cvd2014}, LIVE-VQC \cite{sinno2018large}, LIVE-Qualcomm \cite{ghadiyaram2017capture} and KoNViD-1k \cite{hosu2017konstanz}.}

\textbf{CVD2014.} In this dataset, 78 different cameras, ranging from low-quality phone cameras to dedicated digital single-lens reflex cameras, are used to capture these 234 videos. In particular, five unique scenes (traffic, city, talking head, newspaper and television) are covered with these videos of two resolutions 480P ($640\times480$) and 720P ($1280\times720$).

\textbf{LIVE-VQC.} Videos in this dataset are acquired by 80 inexperienced mobile camera users, leading to a variety of authentic distortions levels. There are in total 585 video scenes in this dataset, containing 18 different resolutions ranging from $1920\times1080$ to $320\times240$.

\textbf{LIVE-Qualcomm.} This dataset consists of 208 videos in total, which are recorded by 8 different mobile phones in 54 different scenes. Six common in-capture distortion categories are studied in this database including: noise and blockiness distortions; incorrect or insufficient color representation; over/under-exposure;  autofocus related distortions; unsharpness and camera shaking. All these sequences have  identical resolution 1080P and quite close frame rate.

\textbf{KoNViD-1k.} KoNViD-1k is the largest VQA dataset which contains in total 1200 video sequences. These videos are sampled from YFCC100m \cite{thomee2016yfcc100m} dataset. Various devices are used to acquire these videos, leading to 12 different resolutions. A portion of the videos in the dataset are acquired by professional photographers, such that there is a large variance in terms of the video quality.

\begin{figure*}[t]
\begin{minipage}[b]{1\linewidth}
  \centering
  \centerline{\includegraphics[width=1\linewidth]{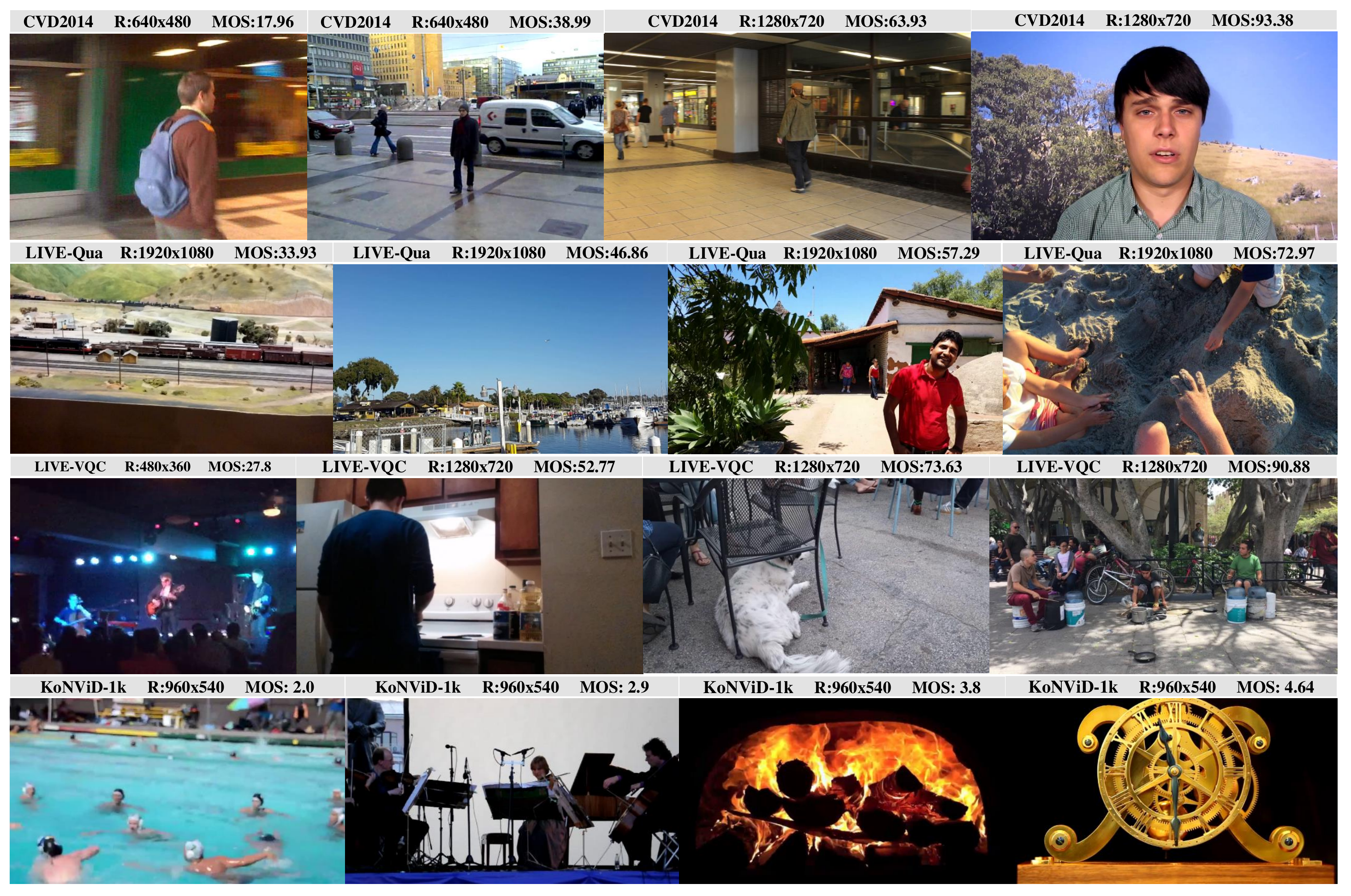}}
\end{minipage}
\caption{Sample frames from four video datasets. The corresponding resolution ($R$) and $MOS$ values are also provided.}
\label{fig:data}
\end{figure*}

In Fig.~\ref{fig:data}, the sampled frames from above four datasets are shown, from which we can observe that these videos are featured by diverse scenes (e.g., indoors and outdoors), resolutions (from $1920\times1080$ to $320\times240$) as well as quality levels. In view of the diverse content, resolutions and frame rates in real-world applications, there has been an exponential increase in the demand for the development of VQA models with high generalization capability.

\begin{figure}[ht]
\begin{center}
\includegraphics[width=0.6\textwidth]{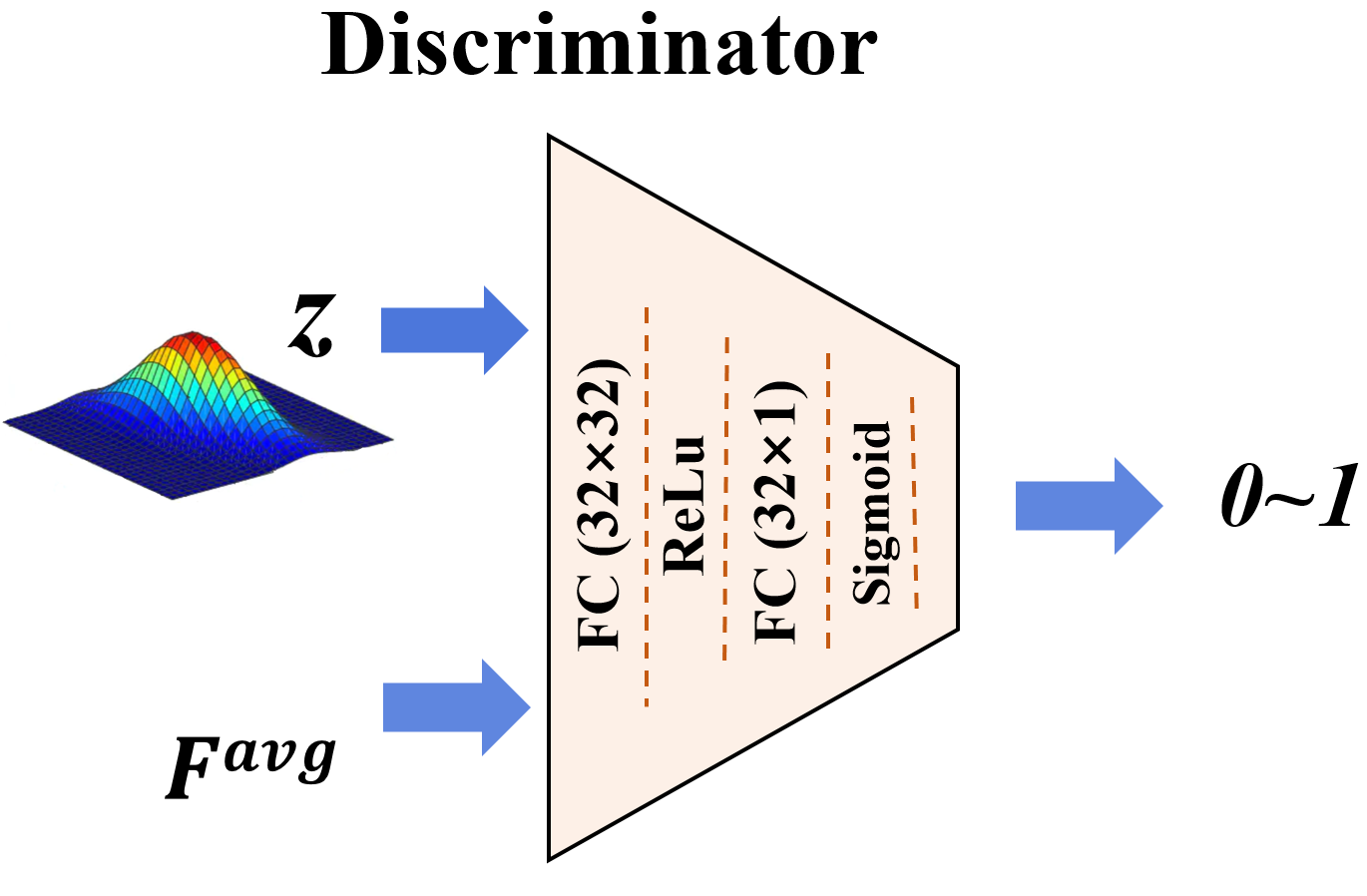}
\caption{The architecture of the proposed discriminator.}
\label{fig:dis}
\end{center}
\end{figure}

\begin{table}
  \centering
  \caption{Architecture of the network in the proposed method.}
\resizebox{230pt}{155pt}{
\begin{tabular}{c|c|ccc}
\hline
\multicolumn{2}{c|}{\textbf{Layer Type}} & \multicolumn{1}{m{3.79em}<{\centering}|}{\textbf{Kernel Size}} & \multicolumn{1}{m{4.79em}<{\centering}|}{\textbf{Channel (in,out)}} & \multicolumn{1}{m{2.09em}<{\centering}}{\textbf{Stride}} \bigstrut\\
\hline
\hline
\multicolumn{1}{c|}{\multirow{9}[18]{*}{\textbf{\shortstack{VGG16 \\  Backbone}}}} & $CNN$×2 & \multicolumn{1}{c|}{3} & \multicolumn{1}{c|}{(3,64)} & 1 \bigstrut\\
\cline{2-5}      & \multicolumn{4}{c}{$MaxPooling$ (stride=2)} \bigstrut\\
\cline{2-5}      & $CNN$×2 & \multicolumn{1}{c|}{3} & \multicolumn{1}{c|}{(64,128)} & 1 \bigstrut\\
\cline{2-5}      & \multicolumn{4}{c}{$MaxPooling$ (stride=2)} \bigstrut\\
\cline{2-5}      & $CNN$×3 & \multicolumn{1}{c|}{3} & \multicolumn{1}{c|}{(128,256)} & 1 \bigstrut\\
\cline{2-5}      & \multicolumn{4}{c}{$MaxPooling$ (stride=2)} \bigstrut\\
\cline{2-5}      & $CNN$×3 & \multicolumn{1}{c|}{3} & \multicolumn{1}{c|}{(256,512)} & 1 \bigstrut\\
\cline{2-5}      & \multicolumn{4}{c}{$MaxPooling$ (stride=2)} \bigstrut\\
\cline{2-5}      & $CNN$×3 & \multicolumn{1}{c|}{3} & \multicolumn{1}{c|}{(256,512)} & 1 \bigstrut\\
\hline
\multicolumn{1}{c|}{\multirow{4}[8]{*}{\textbf{\shortstack{  Attention \\ module}}}} & $FC_1$  &       & (1472,320) &  \bigstrut\\
\cline{2-5}      & \multicolumn{4}{c}{$ReLu$} \bigstrut\\
\cline{2-5}      &$FC_2$  &       & (320,1472) &  \bigstrut\\
\cline{2-5}      & \multicolumn{4}{c}{$Sigmoid$} \bigstrut\\
\hline
\multirow{2}[4]{*}{} & $FC_3$  &       & (1472,256) &  \bigstrut\\
\cline{2-5}      & $GRU$   &       & (256,32) &  \bigstrut\\
\hline
\multicolumn{1}{c|}{\multirow{6}[12]{*}{\textbf{\shortstack{Pyramid\\ Aggregation}}}} & $\textit{1D-}Conv_1$ & \multicolumn{1}{c|}{15} & \multicolumn{1}{c|}{(32,1)} & 1 \bigstrut\\
\cline{2-5}      & \multicolumn{4}{c}{$ReLu$} \bigstrut\\
\cline{2-5}      & $\textit{1D-}Conv_2$ & \multicolumn{1}{c|}{15} & \multicolumn{1}{c|}{(1,1)} & 1 \bigstrut\\
\cline{2-5}      & \multicolumn{4}{c}{$Tahn$} \bigstrut\\
\cline{2-5}      & $FC_4$  &       & (32,1) &  \bigstrut\\
\cline{2-5}      & $FC_5$  &       & (127,1) &  \bigstrut\\
\hline
\hline
\end{tabular}%
}
  \label{tab:network}%
\end{table}%

\begin{table*}
  \centering
  \caption{\bbl{Performance comparisons on four datasets with cross-dataset settings. In each column, the best and second-best values are marked in boldface and underlined, respectively.}}
\resizebox{460pt}{250pt}
{
\begin{tabular}{c|c|cccc|cccc|cccc}
\hline
\multicolumn{2}{c|}{\multirow{2}[4]{*}{\textbf{Training on KoNViD-1k}}} & \multicolumn{4}{c|}{\textbf{CVD2014}} & \multicolumn{4}{c|}{\textbf{LIVE-Qualcomm}} & \multicolumn{4}{c}{\textbf{LIVE-VQC}} \\
\cline{3-14}\multicolumn{2}{c|}{} & \textbf{SROCC} & \textbf{PLCC} & \textbf{PWRC} & \textbf{KROCC} & \textbf{SROCC} & \textbf{PLCC} & \textbf{PWRC} & \textbf{KROCC} & \textbf{SROCC} & \textbf{PLCC} & \textbf{PWRC} & \textbf{KROCC} \\
\hline
\multirow{5}[2]{*}{NR-IQA} & NIQE  & 0.3856  & 0.4410  &  1.7539     & 0.2681  & 0.1807  & 0.1672  &   1.0636    & 0.1196  & 0.4573  & 0.4025  &  1.9544     & 0.3154  \\
      & BRISQUE & 0.4626  & 0.5060  &   2.1754    & 0.3238  & 0.3061  & 0.3303  &    1.4560   & 0.2071  & 0.5805  & 0.5788  &   2.4132    & 0.4089  \\
      & WaDIQaM & \underline{0.6988} & 0.7151 & 3.2439  & \underline{0.5081} & 0.4926 & 0.5471 & 1.9970  & 0.3545 & 0.6461 & 0.6797 & 3.0997  & 0.4634 \\
      & NIMA  & 0.5446 & 0.5836 & 2.4983  & 0.3818 & 0.3413 & 0.4011 & 1.5476  & 0.2003 & 0.5642 & 0.6204 & 2.5660  & 0.3932 \\
      & SPAQ  & 0.6188 & 0.6151 & 2.9039  & 0.4339 & \underline{0.6188} & 0.6151 & 1.0560  & 0.4339 & 0.4653 & 0.5202 & 2.0865  & 0.3202 \\
\hline
\multirow{5}[4]{*}{NR-VQA} & VSFA  & 0.6278  & 0.6216 &  2.9590     & 0.4489 & 0.5574 & 0.5769 &  2.5571     & 0.3966 & 0.6792 & 0.7198  &   3.2374    & \underline{0.4905} \\
      & TLVQM & 0.3569 & 0.3838 & 1.3699  & 0.2442 & 0.4730 & 0.5127 & 2.0319  & 0.3290 & 0.5953 & 0.6248 & 2.7347  & 0.4268 \\
      & VIDEVAL & 0.6494 & 0.6638 & 3.2831  & 0.4684 & 0.4048 & 0.4351 & 1.6906  & 0.2758 & 0.5318 & 0.5329 & 2.3962  & 0.3685 \\
      & CNN-TLVQM  & 0.6828  & \underline{0.7226}  &   \underline{3.3805}    & 0.5003  & 0.6050  & \underline{0.6223}  &    \underline{3.0007}   & \underline{0.4354}  & \textbf{0.7132 } & \textbf{0.7522 } &     \textbf{3.6130}  & \textbf{0.5218 } \\
\cline{2-14}      & \textbf{Ours} & \textbf{0.7972 } & \textbf{0.7984 } & \textbf{4.0625 } & \textbf{0.5891 } & \textbf{0.6200 } & \textbf{0.6666 } & \textbf{3.0198 } & \textbf{0.4445 } & \underline{0.6797 } & \underline{0.7327 } & \underline{3.2126 } & 0.4864  \\
\hline
\multicolumn{1}{r}{} & \multicolumn{1}{r}{} &       &       &       & \multicolumn{1}{r}{} &       &       &       & \multicolumn{1}{r}{} &       &       &       &  \\
\multicolumn{1}{r}{} & \multicolumn{1}{r}{} &       &       &       & \multicolumn{1}{r}{} &       &       &       & \multicolumn{1}{r}{} &       &       &       &  \\
\hline
\multicolumn{2}{c|}{\multirow{2}[4]{*}{\textbf{Training on LIVE-Qualcomm}}} & \multicolumn{4}{c|}{\textbf{KoNViD-1k}} & \multicolumn{4}{c|}{\textbf{CVD2014}} & \multicolumn{4}{c}{\textbf{LIVE-VQC}} \\
\cline{3-14}\multicolumn{2}{c|}{} & \textbf{SROCC} & \textbf{PLCC} & \textbf{PWRC} & \textbf{KROCC} & \textbf{SROCC} & \textbf{PLCC} & \textbf{PWRC} & \textbf{KROCC} & \textbf{SROCC} & \textbf{PLCC} & \textbf{PWRC} & \textbf{KROCC} \\
\hline
\multirow{5}[2]{*}{NR-IQA} & NIQE  & 0.4564  & 0.3619  &  2.2080     & 0.3148  & 0.3856  & 0.4410  &   1.7539    & 0.2681  & 0.4573  & 0.4025  &   1.9544    & 0.3154  \\
      & BRISQUE & 0.4370  & 0.4274  &   2.0893    & 0.2983  & 0.4626  & 0.5060  &    2.1754   & 0.3238  & 0.5805  & 0.5788  &    2.4132   & 0.4089  \\
      & WaDIQaM & 0.3671 & 0.3510 & 1.7712  & 0.2538 & 0.3189 & 0.3255 & 1.1682  & 0.2189 & 0.5385 & 0.5377 & 2.2247  & 0.3756 \\
      & NIMA  & 0.2877 & 0.2588 & 1.3210  & 0.1948 & 0.2705 & 0.2768 & 1.2180  & 0.1842 & 0.3401 & 0.3711 & 1.5301  & 0.2306 \\
      & SPAQ  & 0.1330 & 0.1541 & 0.6490  & 0.0898 & 0.1663  & 0.1508 & 0.8080  & 0.1116 & 0.2854 & 0.3122 & 1.3306  & 01926 \\
\hline
\multirow{5}[4]{*}{NR-VQA} & VSFA  & \underline{0.6643} & \underline{0.6116} &    \underline{3.1701 }    & \underline{0.4769} & 0.5348 & 0.5606 &  2.2852     & 0.3751 & \textbf{0.6425} & \textbf{0.6819 } &    \textbf{3.004 }   & \textbf{0.4613} \\
      & TLVQM & 0.0347 & 0.0467 & 1.4520  & 0.0205 & 0.4893 & 0.4721 & 2.1918  & 0.3361 & 0.4091 & 0.3559 & 1.8241  & 0.2763 \\
      & VIDEVAL & 0.1812 & -0.3441 & 1.0368  & 0.1113  & \underline{0.6059} & \underline{0.6244} & \underline{3.0739 } & \underline{0.4246} & 0.4314 & 0.4122 & 1.8863  & 0.2931 \\
      & CNN-TLVQM  & 0.0854  & 0.0216  &    0.2960   & 0.0692  & 0.2367  & 0.2388  &    0.1906   & 0.1924  & 0.0693  & 0.1040  &  0.0168     & 0.0567  \\
\cline{2-14}      & \textbf{Ours} & \textbf{0.6694 } & \textbf{0.6258 } & \textbf{3.2347 } & \textbf{0.4847 } & \textbf{0.7046 } & \textbf{0.6665 } & \textbf{3.5204 } & \textbf{0.5115 } & \underline{0.6201 } & \underline{0.6100}  & \underline{2.8957 } & \underline{0.4397 } \\
\hline
\multicolumn{1}{r}{} & \multicolumn{1}{r}{} &       &       &       & \multicolumn{1}{r}{} &       &       &       & \multicolumn{1}{r}{} &       &       &       &  \\
\multicolumn{1}{r}{} & \multicolumn{1}{r}{} &       &       &       & \multicolumn{1}{r}{} &       &       &       & \multicolumn{1}{r}{} &       &       &       &  \\
\multicolumn{1}{r}{} & \multicolumn{1}{r}{} &       &       &       & \multicolumn{1}{r}{} &       &       &       & \multicolumn{1}{r}{} &       &       &       &  \\
\hline
\multicolumn{2}{c|}{\multirow{2}[4]{*}{\textbf{Training on LIVE-VQC}}} & \multicolumn{4}{c|}{\textbf{KoNViD-1k}} & \multicolumn{4}{c|}{\textbf{CVD2014}} & \multicolumn{4}{c}{\textbf{LIVE-Qualcomm}} \\
\cline{3-14}\multicolumn{2}{c|}{} & \textbf{SROCC} & \textbf{PLCC} & \textbf{PWRC} & \textbf{KROCC} & \textbf{SROCC} & \textbf{PLCC} & \textbf{PWRC} & \textbf{KROCC} & \textbf{SROCC} & \textbf{PLCC} & \textbf{PWRC} & \textbf{KROCC} \\
\hline
\multirow{5}[2]{*}{NR-IQA} & NIQE  & 0.4564  & 0.3619  &   2.2080    & 0.3148  & 0.3856  & 0.4410  &    1.7539   & 0.2681  & 0.1807  & 0.1672  &   1.0636    & 0.1196  \\
      & BRISQUE & 0.4370  & 0.4274  &   2.0893    & 0.2983  & 0.4626  & 0.5060  &    2.1754   & 0.3238  & 0.3601  & 0.3303  &    1.4560   & 0.2071  \\
      & WaDIQaM & 0.4352 & 0.4451 & 2.0415  & 0.2997 & 0.5362 & 0.5417 & 2.5940  & 0.3666 & 0.4049 & 0.4207 & 1.1650  & 0.2760 \\
      & NIMA  & 0.5848 & 0.5988 & 2.6922  & 0.4105 & 0.3532 & 0.3835 & 1.5811  & 0.2427 & 0.3106 & 0.3362 & 1.3878  & 0.2098 \\
      & SPAQ  & 0.3542 & 0.3468 & 1.6199  & 0.2048 & 0.5494 & 0.4982 & 2.5917  & 0.3837 & 0.2714 & 0.3235 & 1.4808  & 0.1811 \\
\hline
\multirow{5}[4]{*}{NR-VQA} & VSFA  & \underline{0.6584} & \underline{0.6666} & \underline{3.1670} & \underline{0.4751} & 0.5061 & 0.5415 & 2.1028      & 0.3623 & 0.5094 & 0.5350 &  2.3449     & 0.3551 \\
      & TLVQM & 0.6023 & 0.5943 & 2.8976  & 0.4289 & 0.4553 & 0.4749 & 2.0260  & 0.3134 & \underline{0.6415} & \underline{0.6534} & \underline{3.1285 } & \underline{0.4599} \\
      & VIDEVAL & 0.5007 & -0.4841 & 2.3894  & 0.3422 & 0.5702 & 0.5171 & 2.2621  & 0.4125 & 0.3021 & 0.3602 & 1.2132  & 0.2064 \\
      & CNN-TLVQM  & 0.6431  & 0.6304  &   3.253    & 0.4596  & \underline{0.6300 } & \underline{0.6568 } & \underline{2.9609} & \underline{0.4559 } & \textbf{0.6574 } & \textbf{0.6696 } &     \textbf{3.2816}  & \textbf{0.4791 } \\
\cline{2-14}      & \textbf{Ours} & \textbf{0.7085 } & \textbf{0.7074 } & \textbf{3.4437 } & \textbf{0.5179 } & \textbf{0.6894 } & \textbf{0.6645 } & \textbf{3.2897 } & \textbf{0.4888 } & 0.5952  & 0.6245  & 2.9097  & 0.4285  \\
\hline
\multicolumn{1}{r}{} & \multicolumn{1}{r}{} &       &       &       & \multicolumn{1}{r}{} &       &       &       & \multicolumn{1}{r}{} &       &       &       &  \\
\multicolumn{1}{r}{} & \multicolumn{1}{r}{} &       &       &       & \multicolumn{1}{r}{} &       &       &       & \multicolumn{1}{r}{} &       &       &       &  \\
\hline
\multicolumn{2}{c|}{\multirow{2}[4]{*}{\textbf{Training on CVD2014}}} & \multicolumn{4}{c|}{\textbf{KoNViD-1k}} & \multicolumn{4}{c|}{\textbf{LIVE-Qualcomm}} & \multicolumn{4}{c}{\textbf{LIVE-VQC}} \\
\cline{3-14}\multicolumn{2}{c|}{} & \textbf{SROCC} & \textbf{PLCC} & \textbf{PWRC} & \textbf{KROCC} & \textbf{SROCC} & \textbf{PLCC} & \textbf{PWRC} & \textbf{KROCC} & \textbf{SROCC} & \textbf{PLCC} & \textbf{PWRC} & \textbf{KROCC} \\
\hline
\multirow{5}[2]{*}{NR-IQA} & NIQE  & 0.4564  & 0.3619  &   2.2080    & 0.3148  & 0.1807  & 0.1672  &    1.0636   & 0.1196  & 0.4573  & 0.4025  &    1.9544   & 0.3154  \\
      & BRISQUE & 0.4370  & 0.4274  &  2.0893     & 0.2983  & 0.3061  & 0.3303  &   1.4560    & 0.2071  & 0.5805  & 0.5788  &   2.4132    & 0.4089  \\
      & WaDIQaM & 0.4981 & 0.4825 & 2.6027  & 0.3456 & 0.2863 & 0.3305 & 1.1680  & 0.1906 & 0.4598 & 0.5086 & 1.8478  & 0.3222 \\
      & NIMA  & 0.3142 & 0.3013 & 1.5061  & 0.2120 & 0.0294 & 0.0628 & 0.1220  & 0.0189 & 0.2769 & 0.2933 & 1.3025  & 0.1857 \\
      & SPAQ  & 0.3253 & 0.3335 & 1.5053  & 0.2209 & 0.1523 & 0.1951 & 0.6067  & 0.0996 & 0.3619 & 0.4066 & 1.6269  & 0.2482 \\
\hline
\multirow{5}[4]{*}{NR-VQA} & VSFA  &  0.5759 & \underline{0.5636} &     \underline{2.7670}   &  \underline{0.4108} & 0.3256 & 0.3718 &   1.0461    & 0.2192 & 0.4600 & 0.4783 &  1.9002     & 0.3171 \\
      & TLVQM & 0.5437 & 0.5052 & 2.5258  & 0.3758 & 0.3334 & 0.3838 & 1.1622  & 0.2279 & \underline{0.5397} & 0.5527 & \underline{2.4110}  & \underline{0.3803} \\
      & VIDEVAL & 0.1918 & -0.3260 & 1.2059  & 0.1220 & 0.1208 & 0.3315 & 0.5848  & 0.0809 & 0.4751 & 0.5167 & 1.7935  & 0.3192 \\
      & CNN-TLVQM  & \underline{0.5779}  & 0.5489  &   2.6767    & 0.4004  & \textbf{0.4410 } & \underline{0.4712 } &  \textbf{1.8800}     & \textbf{0.2999 } & 0.5209  & \underline{0.5592 } &   2.3276    & 0.3625  \\
\cline{2-14}      & \textbf{Ours} & \textbf{0.6230 } & \bf{0.5764 } & \textbf{3.0248 } & \textbf{0.4437 } & \underline{0.4187 } & \textbf{0.4965 } & \underline{1.8087 } & \underline{0.2857 } & \textbf{0.5817 } & \textbf{0.5751 } & \textbf{2.6587 } & \textbf{0.4090 } \\
\hline
\end{tabular}%
}

  \label{tab:cross}%
\end{table*}%

\subsubsection{Implementation details}
we implement  our  model  by  PyTorch~\cite{paszke2019pytorch}. In Table~\uppercase\expandafter{\romannumeral1} and Fig.~\ref{fig:dis}, we detail the layer-wise network of our proposed method. In particular, we retain the original size of each frame as input without the resizing operation. The VGG-16 network is pretrained on ImageNet~\cite{deng2009imagenet} and we fix its parameters when training. \bl{The batch size in the training phase is 128. In particular, we feed the pre-extracted deep features of the 128 videos to our model in a batch. The model is trained end-to-end with the MOS of each video as the label for regression.} We  adopt Adam optimizer for optimization and the learning rate is fixed to 1e-4. The weighting parameters $ \lambda_{1} $, $ \lambda_{2} $ in Eqn.~\eqref{Eqn:loss} are set as 0.5 and 0.05, respectively. \bl{We train our model in an adversarial manner and the generator and discriminator are learned alternately. In particular, for each batch, we first train the discriminator by maximizing $\mathcal{R}^{\mathrm{gan}}$ loss and maintaining other parts (generator) of our model. Then we maintain the discriminator and update the generator by minimizing the $\mathcal{L}^{\mathrm{vid}}$,  $\mathcal{L}^{\mathrm{reg}}$ and the discriminator loss.}
\bl{In both the cross-dataset and intra-dataset  experiments, we fix the maximum epoch as $200$ and the model learned at the latest $200th$ epoch is used for testing.} For every 20 epochs ($N=20$), we renew the mean and variance of the predefined distribution $g(x)$ in Eqn.~\eqref{Eqn:gan}. It is worth mentioning that all the experimental settings (hyper-parameters and learning strategy) are fixed. \bl{Five evaluation metrics are reported in this paper, including: Spearman’s rank-order correlation coefficient (SROCC), Kendall’s rank-order correlation coefficient (KROCC), Pearson linear correlation coefﬁcient (PLCC), Root mean square error (RMSE) and Perceptually Weighted Rank Correlation (PWRC) \cite{wu2018perceptually}. In particular, the PWRC  rewards the capability of correctly ranking high-quality
images and suppresses the attention toward insensitive rank mistakes, which is confirmed to be more reliable in recommending the perceptually preferred IQA/VQA model.}
As suggested in~\cite{video2000final}, the predicted quality scores \(\hat{s}\) are passed through a nonlinear logistic mapping function before computing PLCC and RMSE,
\begin{equation}
\tilde{s}=\beta_{1}\left(\frac{1}{2}-\frac{1}{\exp \left(\beta_{2}\left(\hat{s}-\beta_{3}\right)\right)}\right)+\beta_{4} \hat{s}+\beta_{5},
\end{equation}
where \(\beta_{1}\)$\sim$\(\beta_{5}\) are regression parameters to be fitted.

\subsection{Quality Prediction Performance}
In this subsection, we evaluate the performance of our method with four different cross-dataset settings to verify the generalization capability. We compare the proposed method with both NR-IQA methods including NIQE~\cite{mittal2012making}, BRISQ~\cite{mittal2012no}, WaDIQaM~\cite{bosse2017deep}, NIMA~\cite{talebi2018nima}, SPAQ~\cite{fang2020perceptual} \bl{and NR-VQA methods including VSFA~\cite{li2019quality}, TLVQM~\cite{korhonen2019two}, VIDEVAL~\cite{tu2020ugc}, and CNN-TLVQM~\cite{korhonen2020blind}}.

In each setting, the models are trained on one dataset and tested on other three datasets. For deep learning based NR-IQA models, we extract two frames per second of each video in the training set and the MOS of the video is treated as the quality score of the extracted frames for model training. The results are shown in Table~\ref{tab:cross}, from which we can find our method can achieve the best performance on all individual cross-dataset settings which reveals the superior generalization ability of our proposed method. Compared with NR-VQA methods, we can observe that the overall performance of NR-IQA methods is not satisfactory as the temporal information is discarded. However, even the VQA based methods cannot achieve very promising performance in such challenging settings. For example, when the method VIDEV trained on LIVE-Qua dataset, the testing result of SROCC is 0.6059 on CVD2014  dataset while it is degraded significantly to 0.1812 on KoNViD-1k dataset which further demonstrates the large domain gap between the two datasets. As shown in Table \ref{tab:cross}, training on CVD2014 dataset and cross-testing on other three datasets is the most challenging setting as only 234 videos and 5 scenes are involved in CVD2014. The limited data  cause the over-fitting problem. However, our method still leads with a large margin over the second-best method VSFA, demonstrating the robustness and promising generalization capability of our method. 

\bl{Moreover, we also provide the performance of our feature regularization module (\textit{a.k.a} output of Eqn. (9)). The results  comparing with the feature aggregation module are shown in Table~\ref{tab:p1}. From the table, we can observe that the average values (SROCC and PLCC) of the feature aggregation module outperform the feature regularization module on all cross-dataset settings, revealing higher generalization capability achieved by the feature aggregation. However, the averaged frame-wise features  can also provide the global quality estimation of the input video.  Therefore, we further fuse the outputs of Eqs. (9) and (13) ($ Q^{reg }$ and $ Q^{vid}$, respectively) with the weights set by $\lambda$  as follows:
\begin{equation}
Q^{{fus }}=\frac{Q^{{vid }}+\lambda Q^{{reg }}}{1+\lambda}
\end{equation}
We set $\lambda$ from 0.0 - 1.8 and the experimental results are shown in Table~\ref{tab:p2}. From the table, we can find the higher overall performance can be achieved when the  $\lambda$ is set around 0.4. This phenomenon demonstrates that the combination of global quality information (acquired by average pooling) can finally benefit the improvement of generalization capability. }

\begin{table*}
  \centering
  \caption{\bl{Performance comparisons on four datasets with cross-dataset settings. The ``Frame'' represents the output of  feature regularization module and  ``Video'' represents the output of  feature aggregation module. The numbers in bold are the best results.}}
\setlength{\tabcolsep}{1.8mm}{
\begin{tabular}{c|c|ccc|c|ccc|c}
\hline
\multicolumn{2}{c|}{\multirow{2}[2]{*}{\textbf{Module Selected}}} & \multicolumn{3}{c|}{\textbf{Tranined on CVD2014}} & \multirow{2}[2]{*}{\textbf{Avg}} & \multicolumn{3}{c|}{\textbf{Tranined on LIVE-Q}} & \multirow{2}[2]{*}{\textbf{Avg}}  \\
\cline{3-5}\cline{7-9}\multicolumn{2}{c|}{} & \textbf{LIVE-Q} & \textbf{LIVE-V} & \textbf{KoNViD-1k} &       & \textbf{CVD2014} & \textbf{LIVE-V} & \textbf{KoNViD-1k} &   \\
\hline
\multirow{2}[2]{*}{\textbf{SROCC}} & Frame & \textbf{0.5220 } & 0.5750  & 0.5063  & 0.5344  & 0.6492  & 0.5555  & 0.5763  & 0.5937   \\
      & Video & 0.4187  & \textbf{0.5817 } & \textbf{0.6230 } & \textbf{0.5411 } & \textbf{0.7046 } & \textbf{0.6201 } & \textbf{0.6694 } & \textbf{0.6647 }  \\
\hline
\multirow{2}[2]{*}{\textbf{PLCC}} & Frame & \textbf{0.5395 } & \textbf{0.5780 } & 0.4819  & 0.5331  & 0.6372  & 0.5897  & 0.5549  & 0.5939   \\
      & Video & 0.4965  & 0.5751  & \textbf{0.5764 } & \textbf{0.5493 } & \textbf{0.6665 } & \textbf{0.6100 } & \textbf{0.6258 } & \textbf{0.6341 }  \\
\hline
\multicolumn{2}{c|}{\multirow{2}[2]{*}{\textbf{Module Selected}}} & \multicolumn{3}{c|}{\textbf{Tranined on LIVE-V}} & \multirow{2}[2]{*}{\textbf{Avg}} & \multicolumn{3}{c|}{\textbf{Tranined on KoNViD-1k}} & \multirow{2}[2]{*}{\textbf{Avg}}  \\
\cline{3-5}\cline{7-9}\multicolumn{2}{c|}{} & \textbf{CVD2014} & \textbf{LIVE-Q} & \textbf{KoNViD-1k} &       & \textbf{CVD2014} & \textbf{LIVE-Q} & \textbf{LIVE-V} &   \\
\hline
\multirow{2}[2]{*}{\textbf{SROCC}} & Frame & 0.6425  & \textbf{0.5983 } & 0.7035  & 0.6481  & 0.7125  & 0.5384  & 0.5997  & 0.6169   \\
      & Video & \textbf{0.6894 } & 0.5952  & \textbf{0.7085 } & \textbf{0.6644 } & \textbf{0.7972 } & \textbf{0.6200 } & \textbf{0.6797 } & \textbf{0.6990 }  \\
\hline
\multirow{2}[2]{*}{\textbf{PLCC}} & Frame & 0.6271  & 0.5998  & 0.6912  & 0.6394  & 0.7337  & 0.6130  & 0.7112  & 0.6860   \\
      & Video & \textbf{0.6645 } & \textbf{0.6245 } & \textbf{0.7074 } & \textbf{0.6655 } & \textbf{0.7984 } & \textbf{0.6666 } & \textbf{0.7327 } & \textbf{0.7326 }  \\
\hline
\end{tabular}%

}
 \label{tab:p1}%
\end{table*}%

\begin{table*}
  \centering
  \caption{\bl{Performance comparisons on four cross-dataset settings with  different weights are set for score combination. The numbers in bold are the best results.}}
\setlength{\tabcolsep}{1.8mm}{
\begin{tabular}{c|ccc|c|ccc|c}
\hline
\multirow{2}[4]{*}{\textbf{$\lambda$}} & \multicolumn{3}{c|}{\textbf{Tranined on CVD2014}} & \multirow{2}[4]{*}{\textbf{Avg}} & \multicolumn{3}{c|}{\textbf{Tranined on LIVE-Q}} & \multirow{2}[4]{*}{\textbf{Avg}}  \\
\cline{2-4}\cline{6-8}      & \textbf{LIVE-Q} & \textbf{LIVE-V} & \textbf{KoNViD-1k} &       & \textbf{CVD2014} & \textbf{LIVE-V} & \textbf{KoNViD-1k} &   \\
\hline
0.0   & 0.4187  & 0.5817  & 0.6230  & 0.5411  & \textbf{0.7046 } & 0.6201  & 0.6694  & \textbf{0.6647 }  \\
0.2   & 0.4342  & 0.5871  & 0.6313  & 0.5509  & 0.7010  & 0.6201  & \textbf{0.6710 } & 0.6640  \\
0.4   & 0.4491  & 0.5895  & \textbf{0.6341 } & 0.5576  & 0.6958  & \textbf{0.6206 } & 0.6689  & 0.6618  \\
0.6   & 0.4605  & 0.5913  & 0.6327  & 0.5615  & 0.6928  & \textbf{0.6206 } & 0.6642  & 0.6592  \\
0.8   & 0.4723  & 0.5923  & 0.6292  & 0.5646  & 0.6887  & 0.6199  & 0.6583  & 0.6556  \\
1.0   & 0.4781  & 0.5924  & 0.6245  & 0.5650  & 0.6868  & 0.6188  & 0.6351  & 0.6469  \\
1.2   & 0.4847  & \textbf{0.5926 } & 0.6189  & 0.5654  & 0.6858  & 0.6182  & 0.6842  & 0.6627  \\
1.4   & 0.4906  & 0.5924  & 0.6134  & \textbf{0.5655 } & 0.6845  & 0.6179  & 0.6435  & 0.6486  \\
1.6   & 0.4963  & 0.5922  & 0.6080  & \textbf{0.5655 } & 0.6833  & 0.6176  & 0.6391  & 0.6467  \\
1.8   & \textbf{0.5016 } & 0.5915  & 0.6030  & 0.5654  & 0.6820  & 0.6170  & 0.6357  & 0.6449   \\
\hline
\hline
\multirow{2}[4]{*}{\textbf{$\lambda$}} & \multicolumn{3}{c|}{\textbf{Tranined on LIVE-V}} & \multirow{2}[4]{*}{\textbf{Avg}} & \multicolumn{3}{c|}{\textbf{Tranined on KoNViD-1k}} & \multirow{2}[4]{*}{\textbf{Avg}}  \\
\cline{2-4}\cline{6-8}      & \textbf{CVD2014} & \textbf{LIVE-Q} & \textbf{KoNViD-1k} &       & \textbf{CVD2014} & \textbf{LIVE-Q} & \textbf{LIVE-V} &   \\
\hline
0.0   & \textbf{0.6894 } & 0.5952  & 0.7085  & 0.6644  & \textbf{0.7972 } & 0.6200  & 0.6797  & 0.6990   \\
0.2   & 0.6849  & 0.6009  & 0.7123  & 0.6660  & 0.7950  & 0.6205  & 0.6830  & 0.6995  \\
0.4   & 0.6809  & 0.6048  & 0.7145  & \textbf{0.6667 } & 0.7934  & \textbf{0.6219 } & 0.6843  & \textbf{0.6999 } \\
0.6   & 0.6766  & 0.6041  & 0.7158  & 0.6655  & 0.7901  & 0.6197  & \textbf{0.6850 } & 0.6983  \\
0.8   & 0.6738  & \textbf{0.6061 } & 0.7166  & 0.6655  & 0.7859  & 0.6193  & 0.6842  & 0.6965  \\
1.0   & 0.6712  & 0.6047  & 0.7167  & 0.6642  & 0.7818  & 0.6165  & 0.6837  & 0.6940  \\
1.2   & 0.6681  & 0.6044  & \textbf{0.7168 } & 0.6631  & 0.7772  & 0.6146  & 0.6834  & 0.6917  \\
1.4   & 0.6662  & 0.6044  & \textbf{0.7168 } & 0.6625  & 0.7737  & 0.6128  & 0.6821  & 0.6895  \\
1.6   & 0.6638  & 0.6046  & 0.7166  & 0.6617  & 0.7709  & 0.6122  & 0.6806  & 0.6879  \\
1.8   & 0.6615  & 0.6050  & 0.7162  & 0.6609  & 0.7692  & 0.6087  & 0.6787  & 0.6855   \\
\hline
\end{tabular}%

}
 \label{tab:p2}%
\end{table*}%

\subsection{Quality Prediction Performance on Intra-dataset}
\bl{In this subsection, to further verify the effectiveness of our method, we evaluate our method on three intra-datasets including  LIVE-Qualcomm, KoNViD-1k and CVD2014. We compare the proposed method with seven existing methods including  NIQE~\cite{mittal2012making}, BRISQ~\cite{mittal2012no},  CORNIA~\cite{ye2012unsupervised}, VIIDEO~\cite{mittal2015completely}, VIDEVAL~\cite{tu2020ugc}, VSFA~\cite{li2019quality},  and CNN-TLVQM~\cite{korhonen2020blind}}. More specifically, for each dataset, 80\% and 20\% data are used for training and testing, respectively. This procedure is repeated 10 times and the mean and standard deviation of performance values are reported in Table~\ref{tab:intra}. \bl{From Table~\ref{tab:intra}, we can observe that our method can achieve the second-best performance overall performance in terms of both the prediction monotonicity (SROCC, KROCC) and the prediction accuracy (PLCC, RMSE).  \bbl{In particular, for the LIVE-Qualcomm dataset and KoNVid-1k dataset, our method achieves the second-best performance which is comparable with the state-of-the-art method CNN-TLVQM, and has a large gain over other methods.}  This phenomenon reveals that our methods can possess the superior generalization capability without much sacrifice of performance on intra-dataset settings.}

\begin{table*}
  \centering
  \caption{ \bbl{Performance comparisons on three VQA datasets with three intra-dataset settings. Mean and standard deviation (std) of the performance values in 10 runs are reported. The overall performance is obtained by weighted-average performance over all three databases, where weights are in proportional to the size of the dataset. In each row, the best and second-best values are marked in boldface and underlined, respectively.}}
\resizebox{520pt}{85pt}{
\begin{tabular}{c|c|ccccccc|c}
\hline
\multicolumn{2}{c|}{\textbf{Method}} & \textbf{NIQE } & \textbf{BRISQUE } & \textbf{CORNIA } & \textbf{VIIDEO } & \textbf{VBLIINDS } & \textbf{VSFA} & \textbf{CNN-TLVQM  } & \textbf{Ours} \bigstrut\\
\hline
\hline
\multirow{4}[2]{*}{\textbf{Overall }} & \textbf{SROCC} & 0.526 (±  0.055) & 0.643 (± 0.059) & 0.591 (± 0.052) & 0.237 (± 0.073) & 0.686 (± 0.035) & 0.771 (± 0.028) & \textbf{0.822} (± 0.025) & \underline{0.811} (± 0.031) \bigstrut[t]\\
      & \textbf{KROCC} & 0.369 (±  0.041) & 0.465 (± 0.047) & 0.423 (± 0.043) & 0.164 (± 0.050) & 0.503 (± 0.032) & 0.582 (± 0.029) & \textbf{0.634} (± 0.024) & \underline{0.620} (± 0.029) \\
      & \textbf{PLCC} & 0.542 (± 0.054) & 0.625 (± 0.053) & 0.595 (± 0.051) & 0.218 (± 0.070) & 0.660 (± 0.037) & 0.762 (± 0.031) & \textbf{0.829} (± 0.021) & \underline{0.817} (± 0.032) \\
      & \textbf{RMSE} & 4.214 (± 0.323) & 3.895 (± 0.380) & 4.139 (± 0.300) & 5.115 (± 0.285) & 3.753 (±  0.365) & 3.074 (± 0.448) & \textbf{2.547} (± 0.273) & \underline{2.832} (± 0.441) \bigstrut[b]\\
\hline
\multirow{4}[2]{*}{\textbf{LIVE-Qualcomm }} & \textbf{SROCC} & 0.463 (± 0.105) & 0.504 (± 0.147) & 0.460 (± 0.130) & 0.127 (± 0.137) & 0.566 (± 0.078) & 0.737 (± 0.045) & \textbf{0.810} (± 0.045) & \underline{0.801} (± 0.053) \bigstrut[t]\\
      & \textbf{KROCC} & 0.328 (± 0.088) & 0.365 (± 0.111) & 0.324 (± 0.104) & 0.082 (± 0.099) & 0.405 (± 0.074) & 0.552 (± 0.047) & \textbf{0.629} (± 0.045) & \underline{0.620} (± 0.052) \\
      & \textbf{PLCC} & 0.464 (± 0.136) & 0.516 (± 0.127) & 0.494 (± 0.133) & -0.001 (± 0.106) & 0.568 (± 0.089) & 0.732 (± 0.036) & \textbf{0.833} (± 0.029) & \underline{0.825} (± 0.043) \\
      & \textbf{RMSE} & 10.858 (± 1.013) & 10.731 (± 1.33) & 10.759 (± 0.939) & 12.308 (± 0.881) & 10.760 (± 1.231) & 8.863 (± 1.042) & \textbf{6.734} (± 0.815) & \underline{7.605} (± 0.935) \bigstrut[b]\\
\hline
\multirow{4}[2]{*}{\textbf{KoNViD-1k }} & \textbf{SROCC} & 0.544 (± 0.040) & 0.654 (± 0.042) & 0.610 (± 0.034) & 0.298 (± 0.052) & 0.695 (± 0.024) & 0.755 (± 0.025) & \textbf{0.816} (± 0.019)   & \underline{0.814} (± 0.026) \bigstrut[t]\\
      & \textbf{KROCC} & 0.379 (± 0.029) & 0.473 (± 0.034) & 0.436 (± 0.029) & 0.207 (± 0.035) & 0.509 (± 0.020) & 0.562 (± 0.022) & \textbf{0.626} (± 0.018) & \underline{0.621} (± 0.027) \\
      & \textbf{PLCC} & 0.546 (± 0.038) & 0.626 (± 0.041) & 0.608 (± 0.032) & 0.303 (± 0.049) & 0.658 (± 0.025) & 0.744 (± 0.029)  & \underline{0.818} (± 0.019)   & \textbf{0.825} (± 0.043) \\
      & \textbf{RMSE} & 0.536 (± 0.010) & 0.507 (± 0.031) & 0.509 (± 0.014) & 0.610 (± 0.012) & 0.483 (± 0.011) & 0.469 (± 0.054) & \textbf{0.358} (± 0.016)  & \underline{0.399} (± 0.020) \bigstrut[b]\\
\hline
\multirow{4}[2]{*}{\textbf{CVD2014}} & \textbf{SROCC} & 0.489 (± 0.091) & 0.709( ± 0.067) & 0.614( ± 0.075) & 0.023( ± 0.122)  & 0.746 (± 0.056) & \textbf{0.880} ( ± 0.030)  & \underline{0.863} (± 0.037) & 0.831 (± 0.052) \bigstrut[t]\\
      & \textbf{KROCC} & 0.358 (± 0.064)  & 0.518 (± 0.060) & 0.441 (± 0.058) & 0.021 (± 0.081) & 0.562 (± 0.057) & \textbf{0.705} (± 0.044) & \underline{0.677} (± 0.038) & 0.657 (± 0.037) \\
      & \textbf{PLCC} & 0.593 (± 0.065) & 0.715 (± 0.048) & 0.618 (± 0.079) & -0.025 (± 0.144) & 0.753 (± 0.053) & \textbf{0.885} (± 0.031) & 0.880 (± 0.025) & 0.844 (± 0.062) \\
      & \textbf{RMSE} & 17.168 (± 1.318) & 15.197 (± 1.325) & 16.871 (± 1.200) & 21.822 (± 1.152) & 14.292 (± 1.413) & \underline{11.287} (± 1.943) & \textbf{10.323 (± 1.134)} &  11.552 (± 2.014) \bigstrut[b]\\
\hline
\end{tabular}%

}
 \label{tab:intra}%
\end{table*}%

\subsection{Ablation Study}
In this subsection, to reveal the functionalities of different modules in the proposed method, we perform the ablation analysis. The experiments are conducted with a cross-dataset setting (training on KoNViD-1k and testing on other three datasets). As shown in Table~\ref{tab:abl}, the performance are provided in terms of SROCC and PLCC.  To identify the effectiveness of the  attention module used in multi-scale features extraction, we directly concatenate the mean and std pooling features without attention performed and maintain the rest of parts for training. The model is denoted as \textbf{Concat} in Table~\ref{tab:abl}, in which we can observe that the performance on all testing sets is degraded especially on the LIVE-Qualcomm dataset. The similar phenomenon can be observed when the pyramid poling module is ablated (denoted as \textbf{Ours $w/o$ PymidPooling} in Table~\ref{tab:abl}). The reason lies in that the videos in LIVE-Qualcomm dataset challenge both human subjects and objective VQA models, as indicated in \cite{ghadiyaram2017capture}. As such, more dedicated design on both spatial and temporal domains is desired. Moreover, we remove the Gaussian distribution regularization module from the original models, leading to a model denoted as \textbf{Ours $w/o$  Distribution}. From the results, we can find that both the SROCC and PLCC are degraded compared with our original method (denoted as \textbf{Ours}) which demonstrates that the regularization on feature space is also important for the generalized VQA model. 

\begin{table}
  \centering
  \caption{Ablation studies with the KoNViD-1k as the training data. For simplification, we denote the LIVE-Qualcomm as LIVE-Q and LIVE-VQC and LIVE-V.}
\small
{
\begin{tabular}{c|c|ccc}
\hline
\multicolumn{2}{c|}{\textbf{Method}}& \textbf{CVD2014} & \textbf{LIVE-Q} & \textbf{LIVE-V} \bigstrut\\
\hline
\multirow{2}[2]{*}{\textbf{Concat}} & \textbf{SROCC} & 0.7466  & 0.5286  & 0.6357 \bigstrut[t]\\
      & \textbf{PLCC} & 0.7625  & 0.5847  & 0.6889 \bigstrut[b]\\
\hline

\multirow{2}[2]{*}{\textbf{\shortstack{Ours $w/o$\\ Distribution}}} & \textbf{SROCC} & 0.7735  & 0.5900  & 0.6692 \bigstrut[t]\\
      & \textbf{PLCC} & 0.7638  & 0.6524  & 0.7142 \bigstrut[b]\\
\hline
\multirow{2}[2]{*}{\textbf{\shortstack{Ours $w/o$\\ PyramidPooling}}} & \textbf{SROCC} & 0.7732  & 0.5884  & 0.6701 \bigstrut[t]\\
      & \textbf{PLCC} & 0.7631  & 0.6495  & 0.7173 \bigstrut[b]\\
\hline
\multirow{2}[2]{*}{\textbf{Ours}} & \textbf{SROCC} & \textbf{0.7972} & \textbf{0.6200} & \textbf{0.6797} \bigstrut[t]\\
      & \textbf{PLCC} & \textbf{0.7984} & \textbf{0.6666} & \textbf{0.7327} \bigstrut[b]\\
\hline
\end{tabular}%
}
  \label{tab:abl}%
\end{table}%

\subsection{Visualization}
\begin{figure*}[t]
\begin{minipage}[b]{1\linewidth}
  \centering
  \centerline{\includegraphics[width=1\linewidth]{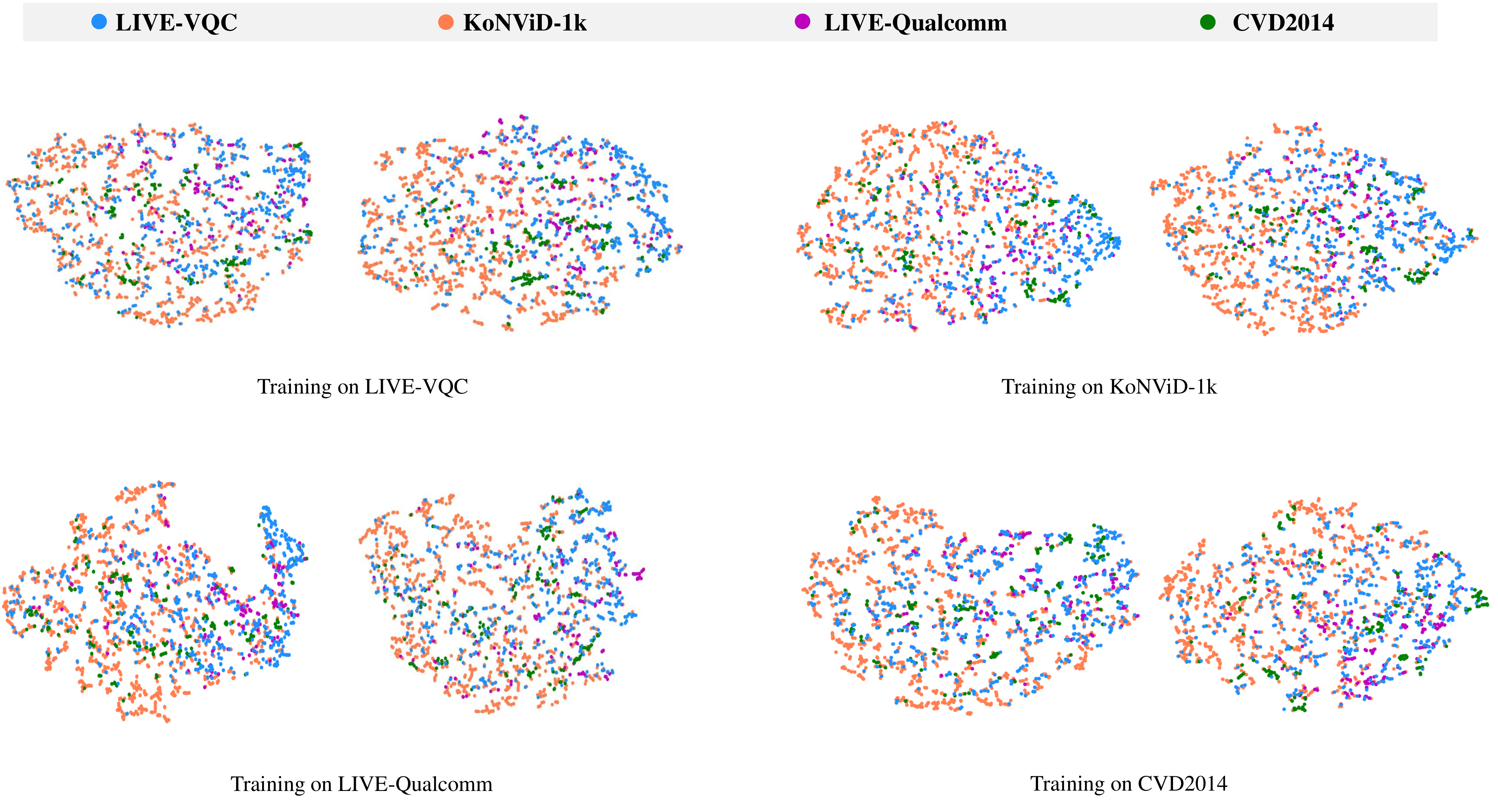}}
\end{minipage}
\caption{T-SNE visualization of the features extractions from each dataset. 
The dataset used for training is provided under each sub-figure. }
\label{fig:tsne}
\end{figure*}

 \bl{To better understand the  quality relevant features learned in our proposed method, we train our model on one specific dataset and visualize the quality features of all videos of the four datasets. More specifically, for each video, we first extract its feature $\textit{F}^{avg}$ (as shown in Eqn.~\eqref{Eqn:gan}) generated with/without the Gaussian distribution based regularization, respectively. Subsequently, the feature dimension is reduced to two by T-SNE~\cite{maaten2008visualizing}, as visualized in Fig.~\ref{fig:tsne}. We are particularly interested in the KoNViD-1k dataset due to the large number of videos with a wide-range of quality levels. As shown in Fig.~\ref{fig:tsne}, we can find the features of KoNViD-1k dataset and the features of other three datastes are more compact when they are regularized. More specifically, when the models are trained on the CVD2014 dataset, a larger domain gap can be observed between the KoNViD-1k dataset and the other three datasets when compared with the model trained with regularization, further verifying the effectiveness of our Gaussian distribution based regularization module.}

\begin{figure*}[t]
\begin{minipage}[b]{1\linewidth}
  \centering
  \centerline{\includegraphics[width=1\linewidth]{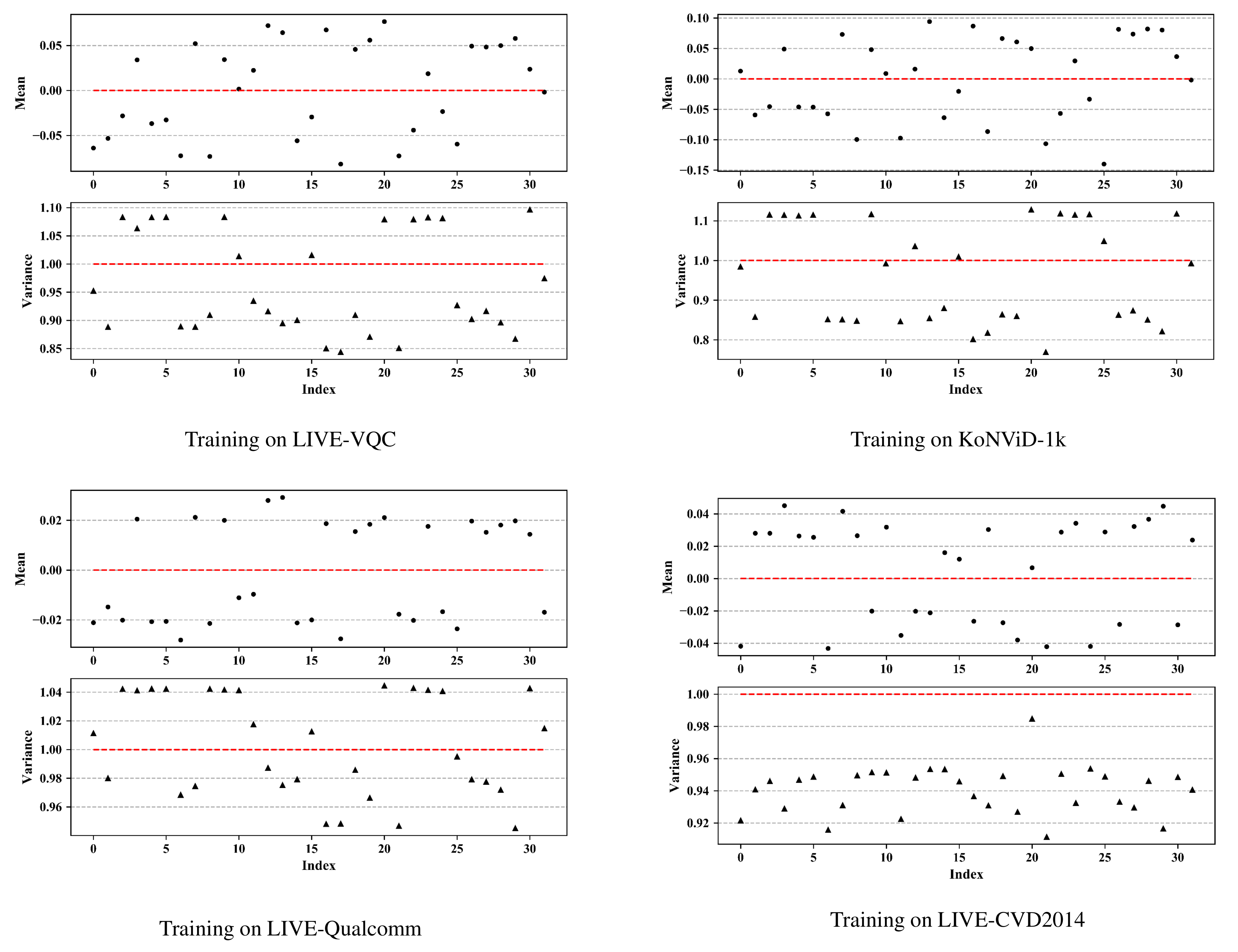}}
\end{minipage}
\caption{{Mean and variance of each dimension of $\textit{F}^{avg}$ in Eqn.~\eqref{Eqn:gan}. The dataset used for training is provided under each sub-figure.}}
\label{fig:mv}
\end{figure*}

Moreover, to verify whether the Gaussian distribution is updated from the initial standard distribution (mean \textit{$\mu=0$} and variance \textit{$\sigma=1$}) of each dimension in $\textit{F}^{avg}$, we also plot the final values of mean and variance in Fig.~\ref{fig:mv} on four cross-dataset testings. We can observe that the distributions of each feature dimension is totally different from each other. For example, when the model is trained on LIVE-VQC dataset, the variance of 30-th dimension is nearly 1.4 times of the 17-th dimension, which further reveals that the quality of the video is governed by the features from different dimensions with different sensitives. 

\section{Conclusions}
In this paper, we propose an NR-VQA method, aiming for improving the generalization capability of the quality assessment model when the training and testing videos hold different content, resolutions and frame rates. The effectiveness of the proposed method, which has been validated in both cross-dataset and intra-dataset settings, arises from the feature learning based upon unified distribution constraint and pyramid temporal aggregation. 
The proposed model is extensible from multiple perspectives. For example, the proposed model can be further applied in the optimization tasks when the pristine reference video is not available. Moreover, the design philosophy could be further applied to other domains (e.g., high dynamic range, screen content, virtual reality).

\bibliographystyle{IEEEtran}
\bibliography{VQACross}

\begin{IEEEbiography}[{\includegraphics[width=1in,height=1.25in,clip,keepaspectratio]{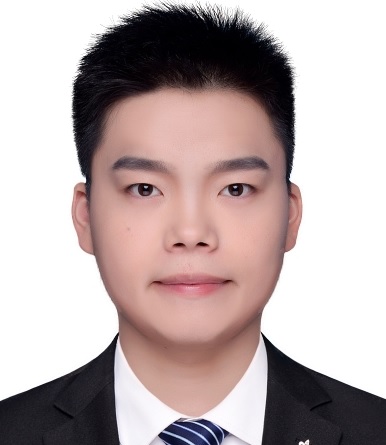}}]{Baoliang Chen} received the B.S. degree in Electronic Information Science and Technology from Hefei University of Technology, Hefei, China, in 2015 and the M.S. degree in Intelligent Information Processing from Xidian University, Xian, China, in 2018. He was a researcher in iFlytek Inc., from 2018 to 2019. He is currently pursuing the Ph.D. degree in Department of Computer Science of City University of HongKong, HongKong. His  research interests include  image/video quality assessment and information security.
\end{IEEEbiography}

\begin{IEEEbiography}[{\includegraphics[width=1in,height=1.25in,clip,keepaspectratio]{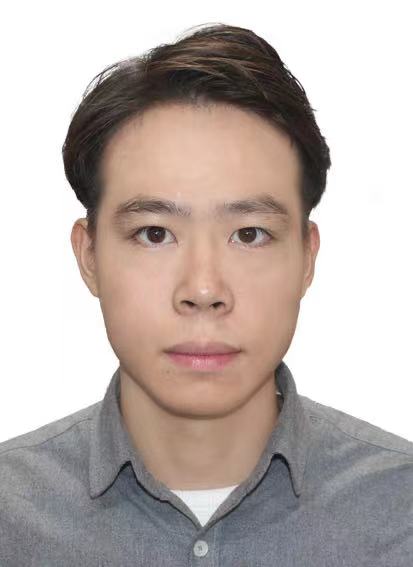}}]{Lingyu Zhu} received the B.S. degree from the Wuhan University of Technology in 2018 and the master's degree from Hong Kong University of Science and Technology in 2019. He is currently pursuing the Ph.D. degree at the City University of Hong Kong. His research interests include image/video quality assessment, image/ video processing, and deep learning.
\end{IEEEbiography}

\begin{IEEEbiography}[{\includegraphics[width=1in,height=1.25in,clip,keepaspectratio]{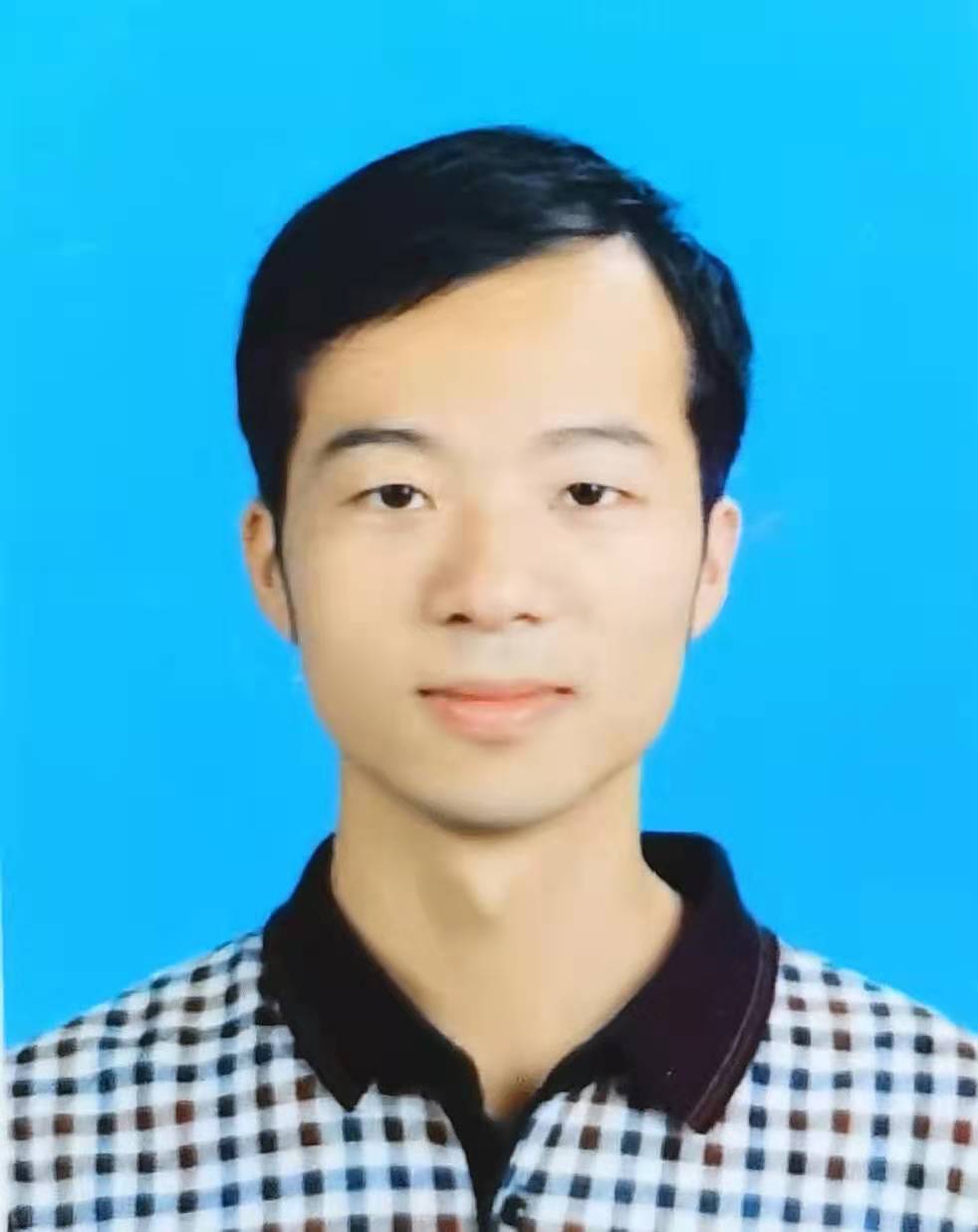}}]{Guo Li} received the B.S. degree in physics from Huazhong University of Science and Technology in 2012 and the M.S. degree in physics from Huazhong University of Science and Technology in 2016. He worked successively in Yuanfudao Company and Kingsoft Cloud Company, China. His research fields include image processing and quality assessment.
\end{IEEEbiography}

\begin{IEEEbiography}[{\includegraphics[width=1in,height=1.25in,clip,keepaspectratio]{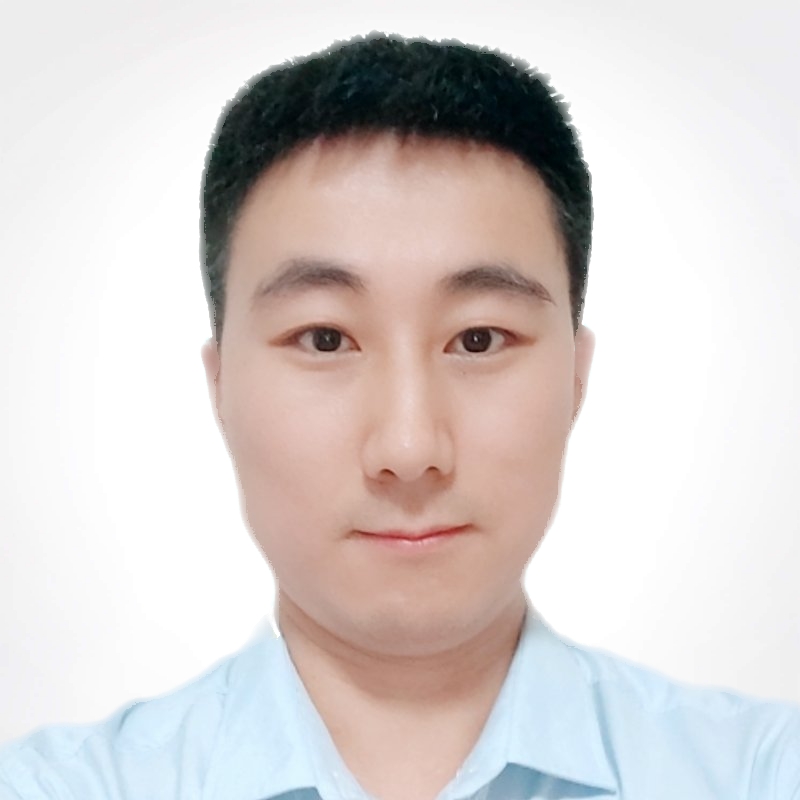}}]{Fangbo Lu} is currently  working on Kingsoft Cloud Company as a Senior Algorithm Engineer, China. His research interests include video processing, video or audio codec.
\end{IEEEbiography}

\begin{IEEEbiography}[{\includegraphics[width=1in,height=1.25in,clip,keepaspectratio]{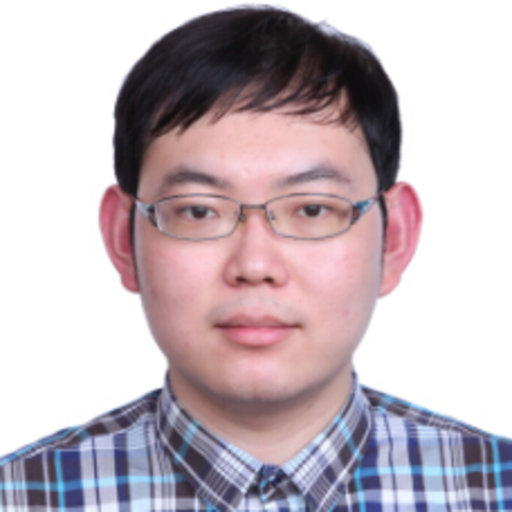}}]{Hongfei Fan} received the B.S. degree in software engineering from Shanghai Jiao Tong University in 2013 and the Ph.D. degree in computer application technology from Peking University in 2017. Since July 2017, he has been working as an Algorithm Architect with Kingsoft Cloud Company, China. His research interests include video coding
and image processing.
\end{IEEEbiography}

\begin{IEEEbiography}[{\includegraphics[width=1in,height=1.25in,clip,keepaspectratio]{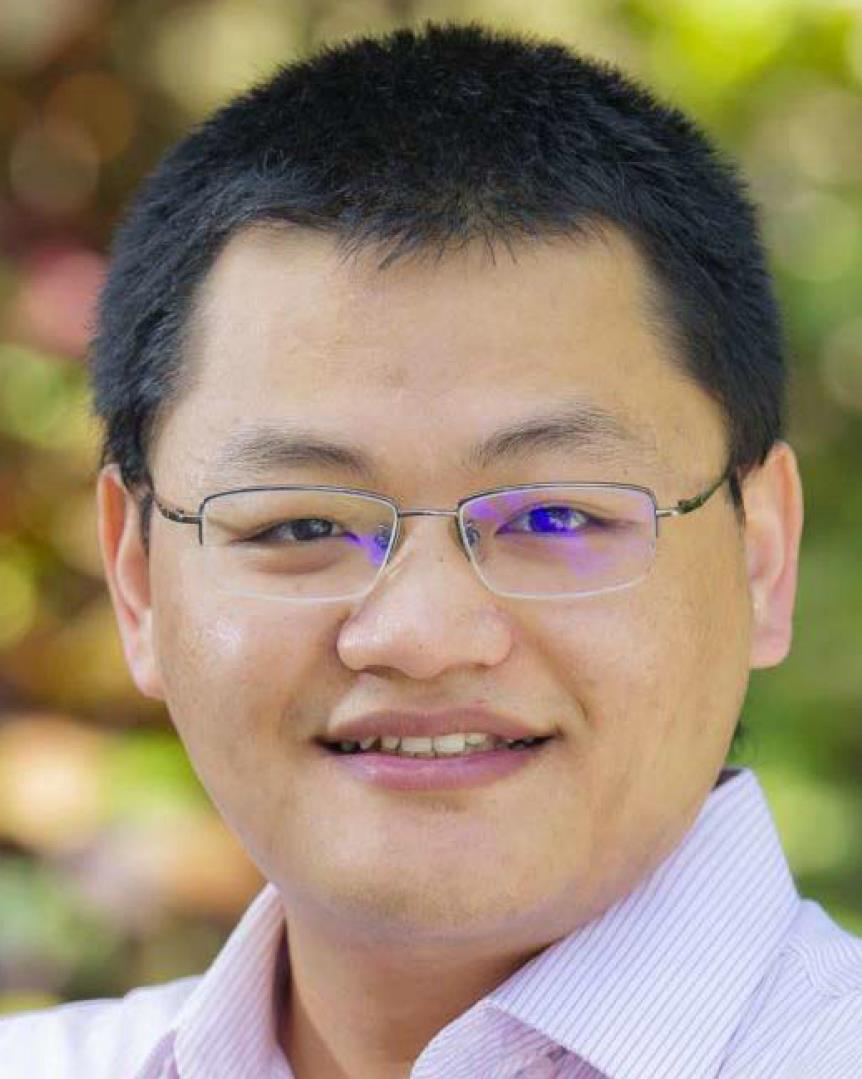}}]{Shiqi Wang}  (Member, IEEE) received the B.S. degree in computer science from the Harbin Institute of Technology in 2008 and the Ph.D. degree in computer application technology from Peking University in 2014. From 2014 to 2016, he was a Post-Doctoral Fellow with the Department of Electrical and Computer Engineering, University of
Waterloo, Waterloo, ON, Canada. From 2016 to
2017, he was a Research Fellow with the Rapid-Rich
Object Search Laboratory, Nanyang Technological
University, Singapore. He is currently an Assistant Professor with the Department of Computer Science, City University of Hong Kong. He has proposed over 50 technical proposals to ISO/MPEG, ITU-T, and AVS standards, and authored/coauthored more than 200 refereed journal articles/conference papers. He received the Best Paper Award from IEEE VCIP 2019, ICME 2019, IEEE Multimedia 2018, and PCM 2017 and is the coauthor of an article that received the Best Student Paper Award in the IEEE ICIP 2018. His research interests include video compression, image/video quality assessment, and image/video search and analysis. 
\end{IEEEbiography}


\end{document}